\documentclass[reprint,amsmath,amssymb,aps,prb,floatfix,superscriptaddress]{revtex4-2}
\usepackage{graphicx}
\usepackage{dcolumn}
\usepackage{bm}
\usepackage{subfigure}
\usepackage{color}
\usepackage{physics}
\usepackage{hyperref}

\begin{document}
\preprint{APS/123-QED}
\title{Structural Distortions and Ferroelectricity in Antiperovskite Oxides with Tetrel Elements}

\author{He Zhu}
\email{zhu00336@umn.edu}
\author{Turan Birol}
\email{tbirol@umn.edu}
\affiliation{Department of Chemical Engineering and Materials Science, University of Minnesota, Minneapolis, Minnesota 55455, USA}
\date{\today}

\begin{abstract}
Antiperovskites share the same structure as perovskites, but allow completely different chemistries and nominal charge states of anions to be stabilized. This gives rise to many interesting phenomena, including septet superconductivity and topological crystalline insulating phases in these systems. Despite this, the work on the crystal structural trends in these compounds is more limited compared to perovskites. In this study, we consider the family of antiperovskite oxides with tetrel elements (Si, Ge, Sn, Pb) and alkaline earth metals (Ca, Sr, Ba), and perform a detailed study of their crystal structures using first principles density functional theory. We show how tolerance factor arguments can be constructed to predict their structure in a way parallel to the perovskites, and furthermore, how heterostructuring (or cation-order) can be used to induce ferroelectricity in these systems which may provide an experimental knob to modify electronic structure. We conclude with a discussion of the electronic structure of antiperovskites, and show that they display interesting trends not observed in regular perovskites, including significant antibonding interactions between face-center ions, which might need to be taken into account when building effective electronic models of these compounds. 
\end{abstract}
\maketitle

\section{\label{sec:1}Introduction}

Inorganic ABX$_3$ perovskites (where A and B are metals; X are pnictogen, chalcogen, or halogen ions) exhibit a great number of exciting and useful properties for applications, as well as provide testbeds for many phenomena of academic interest \cite{pena2001chemical,moure2015recent,manser2016intriguing}. This is in no small part due to the perovskite structure's versatility in chemical composition and structural distortions  \cite{glazer1972classification,woodward1997octahedralI,woodward1997octahedralII}. As the demand for electronic, magnetic, and optical devices increases with an exponential rate, both applied and theoretical work on perovskites have been rapidly increasing in the recent decades.

\begin{figure}[bht]
\includegraphics[width=0.9\linewidth]{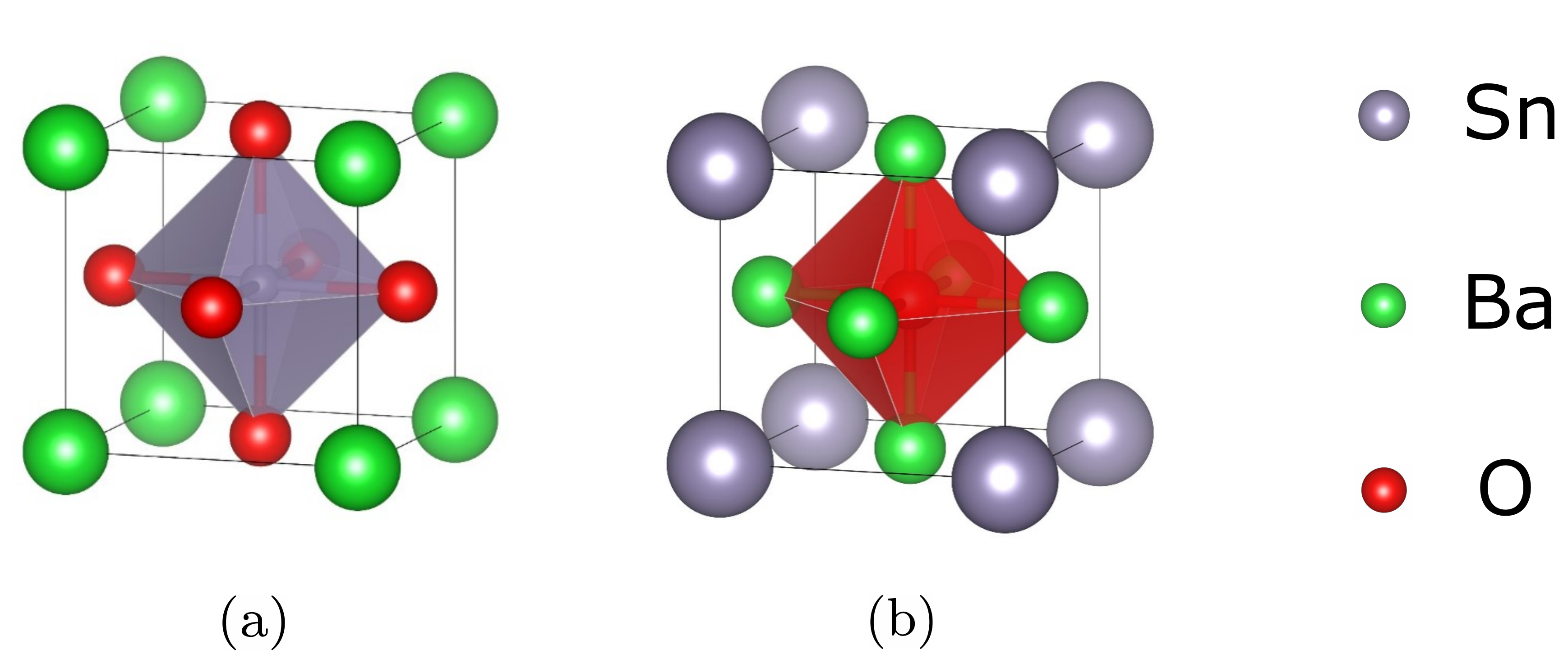}
\caption{\label{fig:perovskite_structure}The crystal structures of cubic (a) perovskite, BaSnO$_3$, and (b) AP, Ba$_3$SnO.}
\end{figure}

It is also possible to chemically stabilize X$_3$AB antiperovskites (AP) with two anions A and B, and a cation X on the face-center sites, effectively inverting the charges of ions in the perovskite structure. (Fig.\ref{fig:perovskite_structure})
Just like their perovskite counterparts, APs possess fascinating properties relevant to various fields such as superconductivity \cite{he2001superconductivity, oudah2016superconductivity, oudah2019evolution, azouaoui2024superconductivity, iyo2021antiperovskite}, topological crystalline insulating phases \cite{hsieh2014topological, chiu2017type, kawakami2018topological}, and multiferroicity \cite{markov2021ferroelectricity}. Even though the volume of research on APs is still far behind that on perovskites, the structure of APs provides the same versatility, and hence, these materials are likely to be a fertile playground for more exciting findings. Thanks to modern growth techniques, an increasing number of new or higher quality single crystals and thin-films of APs are successfully synthesized every year \cite{minohara2018growth, huang2019unusual, wu2021molecular, okamoto2016thermoelectric, pohls2019thermoelectric, sebastia2023chalcohalide}.

Ferroelectricity (the emergence of a spontaneous and switchable electric polarization in solids) is a phenomenon that is studied extensively in perovskites \cite{cohen1992origin,dawber2005physics,rabe2007physics}. In addition to existing ferroelectric perovskites already in use in modern electronic devices \cite{xiong2014active, neelakandan2025review}, novel and higher performance perovskite and perovskite-like ferroelectrics are still being discovered. For example, hybrid improper ferroelectrics (HIF) raised both theoretical and experimental interest recently \cite{benedek2011hybrid, benedek2012polar, harris2011symmetry, zhang2016hybrid, yoshida2018hybrid, xu_highly_2020, smith2021revealing}. Unlike proper ferroelectric perovskites (e.g., BaTiO$_3$), which rely on polar distortions that are driven by negative force constant eigenvalues (dynamical instabilities) at the Brillouin zone center, HIFs result from the interplay of structural instabilities corresponding to zone-boundary phonon modes \cite{benedek2012polar, bousquet2008improper, rondinelli2012octahedral}. The most commonly studied classes of HIFs are the Ruddlesden-Popper phases (e.g., Sr$_3$Ti$_2$O$_7$, Ca$_3$Mn$_2$O$_7$ \cite{liu2015hybrid}) and double perovskites, i.e. short-period heterostructures (e.g., Ba$_2$BiVO$_6$ \cite{parida2019ferroelectric}, CaMnTi$_2$O$_6$ \cite{gou2017site}). 

Given that the parent phase of APs is geometrically the same as the perovskite structure, it is reasonable to propose that HIF also be observed in AP-like structures—double antiperovskites (DAP) or anti-Ruddlesden-Popper (ARP) phases. Some recent studies
discuss the structural properties of APs \cite{wang2020antiperovskites, garcia2020octahedral, mahmood2018investigations, pham2018computational, haddadi2010inverse, stoiber2019perovskite, kaur2020first}, DAPs \cite{Asghar2025, Goh2020, Han2021, Rani2021, mi2022rock}, and ferroelectricity in a DAP \cite{garcia2019hybrid} to different extents. Nevertheless, a comprehensive theory study that compares and contrasts trends in a large number of materials and aims to elucidate ferroelectricity-related properties of all tetrel oxide AP materials is missing to date.

In this study, we employ first-principles calculations to systematically study the emergence of octahedral rotations in AP oxides with tetrel elements, X$_3$TtO (X = Ca, Sr, Ba; Tt = Si, Ge, Sn, Pb), and confirm how the Goldschmidt tolerance factor also applies to these materials. We then show that there are significant differences between perovskites and APs because of the partial covalent nature of bonds in APs. By examining the Landau free energy expressions and coefficients, as obtained from group theory and density functional theory respectively, we explain the absence of the intermediate orthorhombic phase (with $a^0a^0c^-$ octahedral rotations) in the temperature phase diagram of APs. Lastly, we show that HIFs exist in layered double antiperovskites with A-site anion ordering. 

In addition to reproducing and explaining the experimentally observed structures in tetrel oxide APs, our study illustrates that these materials can be driven to polar and in-principle switchable ferroelectric phases by heterostructuring or cation ordering, thus opening up new venues for manipulating their electronic properties of interest.

\section{\label{sec:2}Computational Methods}
Kohn-Sham DFT \cite{kohn1965self, hohenberg1964inhomogeneous} calculations are performed as implemented in the Vienna \textit{Ab initio} Simulation Package (\texttt{VASP}) within the PBEsol generalized gradient approximation adapted for solids \cite{Kresse1996_DFT1, Kresse1996_DFT2, perdew2008restoring}. For structural relaxations of APs and DAPs, 40-atom supercells are constructed with $8\times8\times8$ and $4\times4\times4$ $\Gamma$ centered k-point grids, respectively. 
%
%
Ionic positions, as well as shapes and volume of the unit cells are relaxed until the norms of all the forces are smaller than $0.5\times10^{-3} \mathrm{eV/\AA}$. 

Band structure plots are obtained by using the Heyd-Scuseria-Ernzerhof hybrid functional \cite{heyd2003hybrid}, which captures the bandgap more accurately, but this functional was not used in other (structural relaxation, etc.) calculations because of its computational cost.

Phonon dispersion relations are obtained utilizing the  \texttt{Phonopy} package, which enables calculation of phonon-related properties \cite{togo2015first}. The force constants are generated by the finite displacement method, performing self-consistent DFT calculations only. While the supercell sizes in these calculations limit the resolution of the phonon frequencies at arbitrary momenta, the instabilities discussed in this study are all commensurate with the supercells used, and hence the results on their frequencies are reliable. 

The allowed terms in Landau Free Energy (LFE) expressions were determined using the \texttt{INVARIANTS} tool \cite{hatch2003invariants}. \texttt{INVARIANTS} is an online program that generates LFEs in terms of order parameters represented by the irreducible representations (irreps) for a given space group. The structural distortions relevant to a phase transition correspond to irreps of a given space group, and hence, the structural free energy can be represented as a LFE in terms of them. \texttt{INVARIANTS} generates free energies in terms of polynomials of order parameters defined by the irrep, where all the polynomials transform like the fully symmetric (trivial) representation, i.e. invariant under any symmetry operation.

Covalent bonding is investigated based on Crystal orbital Hamilton population (COHP) analysis. In terms of orbital-pair contributions, COHP partitions the electronic structure as a function of energy into bonding, non-bonding, and anti-bonding regions \cite{deringer2011crystal}. The results of COHP are obtained using \texttt{VASP} and \texttt{LOBSTER} \cite{deringer2011crystal, nelson2020lobster}, a program that calculates projected COHP through post-processing of DFT wavefunctions. The basis functions in \texttt{pbeVaspFit2015} basis set \cite{maintz2016lobster} are chosen as follows: Ca's 4s, 3p, and 3s; Sr's 5s, 4p, 3d, and 4s; Ba's 6s, 5p, 4d, and 5s; Si's 3p, 3s, 2p, and 2s; Ge's 4p, 3d, 3p, and 4s; Sn's 5p, 5s, 4d, and 4p; Pb's 6p, 5d, 4f, and 6s; O's 2p and 2s. In these calculations, primitive unit cells are used. We underline that because different basis wavefunctions are used for each element, comparison of COHP values between different  compounds with different atoms is not always well-defined, and our results should be considered as qualitative results that demonstrate basic trends rather than precise quantitative results.

\section{Results and Discussions}
\subsection{Octahedral Rotations in Antiperovskite Oxides}

\begin{figure}[bht]
\includegraphics[height = 5.0cm]{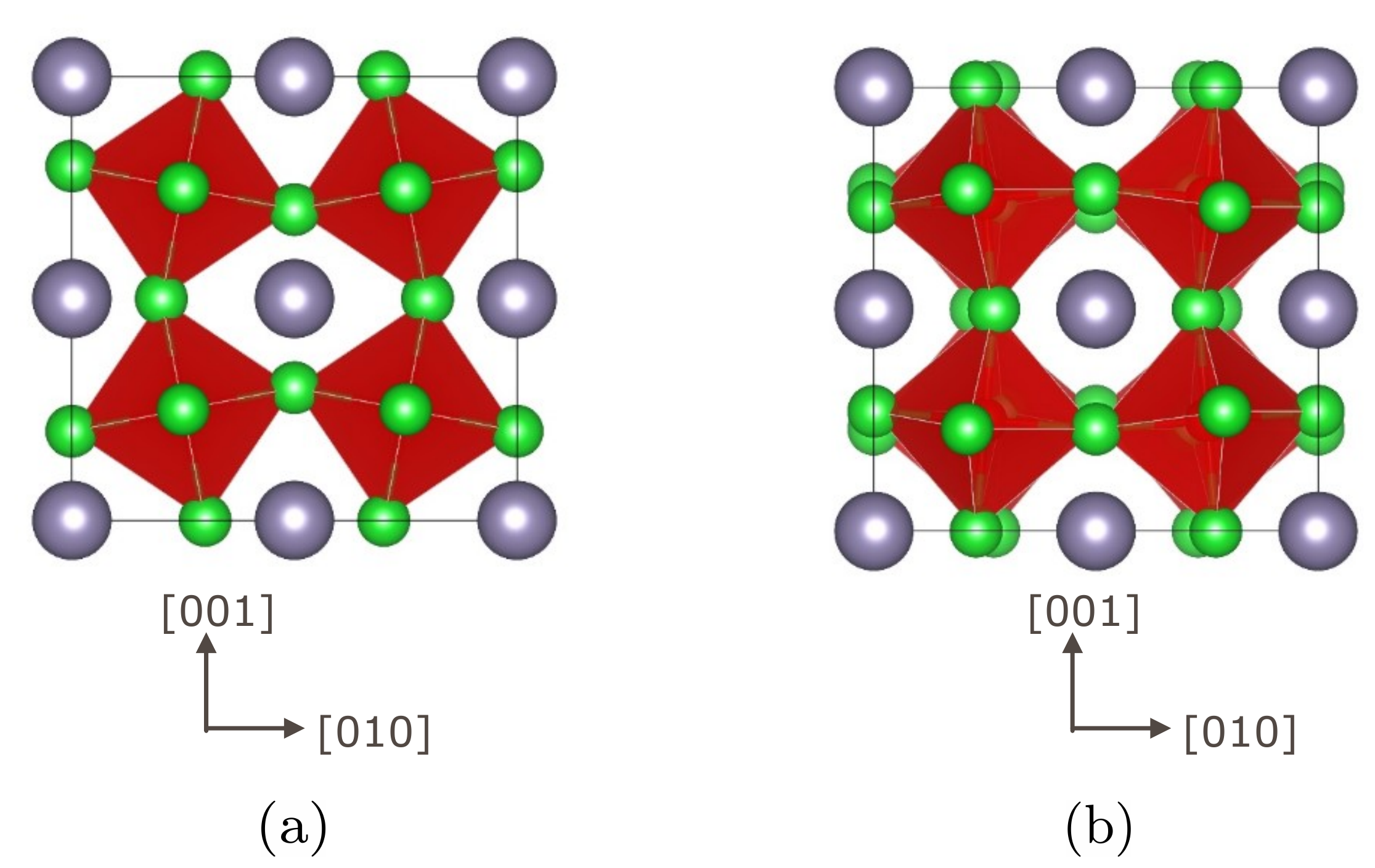}
\caption{\label{fig:rot}Octahedral rotations in perovskites: (a) $a^+b^0b^0$ and (b) $a^-a^-c^+$ rotation patterns.} 
\end{figure}

The octahedral rotations are common in perovskite and perovskite-like structures, and most perovskites don't have cubic symmetry but because of symmetry-lowering octahedral rotations \cite{Lufaso2001}. The A-site ions are usually underbonded, and the rotations help strengthen A-X bonding \cite{woodward1997octahedralII}. Rotations around different axes can coexist in a crystal, and the so-called Glazer notation is widely adopted to describe the rotation pattern, as well as to uniquely determine the space group \cite{glazer1972classification}.
For example, $a^+b^0b^0$ represents a single in-phase rotation around a axis, as shown in Fig. \ref{fig:rot}(a); $a^-a^-c^+$ represents a combination of out-of-phase rotations around both a and b axes, and an in-phase rotation around c axis, as shown in Fig. \ref{fig:rot}(b).

The Goldschmidt tolerance factor 
\begin{equation}
    t_p=\frac{r_A + r_X}{\sqrt{2}\left(r_B+r_X\right)}
\end{equation} 
where $r_i$ represents the ionic radius of atom $i$, is an estimate of the tendency for perovskites to distort from the cubic phase. Typically, smaller tolerance factors are correlated with larger octahedral rotation angles in perovskites. In APs, with the same reference crystal structure, it is natural to ask whether a tolerance factor plays a similar role. In an AP compound $X_3TtO$ with a tetrel element $Tt$, the tolerance factor $t_a$ is defined as $t_a = \left(r_{Tt}+r_X\right)/\sqrt{2}\left(r_O+r_X\right)$, where $r_{Tt}$, $r_{X}$, and $r_{O}$ are ionic radii of $Tt^{-4}$, $X^{-2}$ and $O^{-2}$ respectively. The ionic radii in these formulas are determined by fitting values to bond lengths in large datasets from  different compounds \cite{shannon1976revised}, however, due to the scarcity of $Tt^{-4}$ anions in existing compounds, their radii need to be estimated only using the information about the  $XX'Tt$ phases \cite{bruzzone1978occurrence, liu2006synthesis}. We used the values of $r_{O}$ and $r_{X}$ from \cite{shannon1976revised}. 

\begin{figure}[b]
\subfigure[]{\includegraphics[width=.45\textwidth]{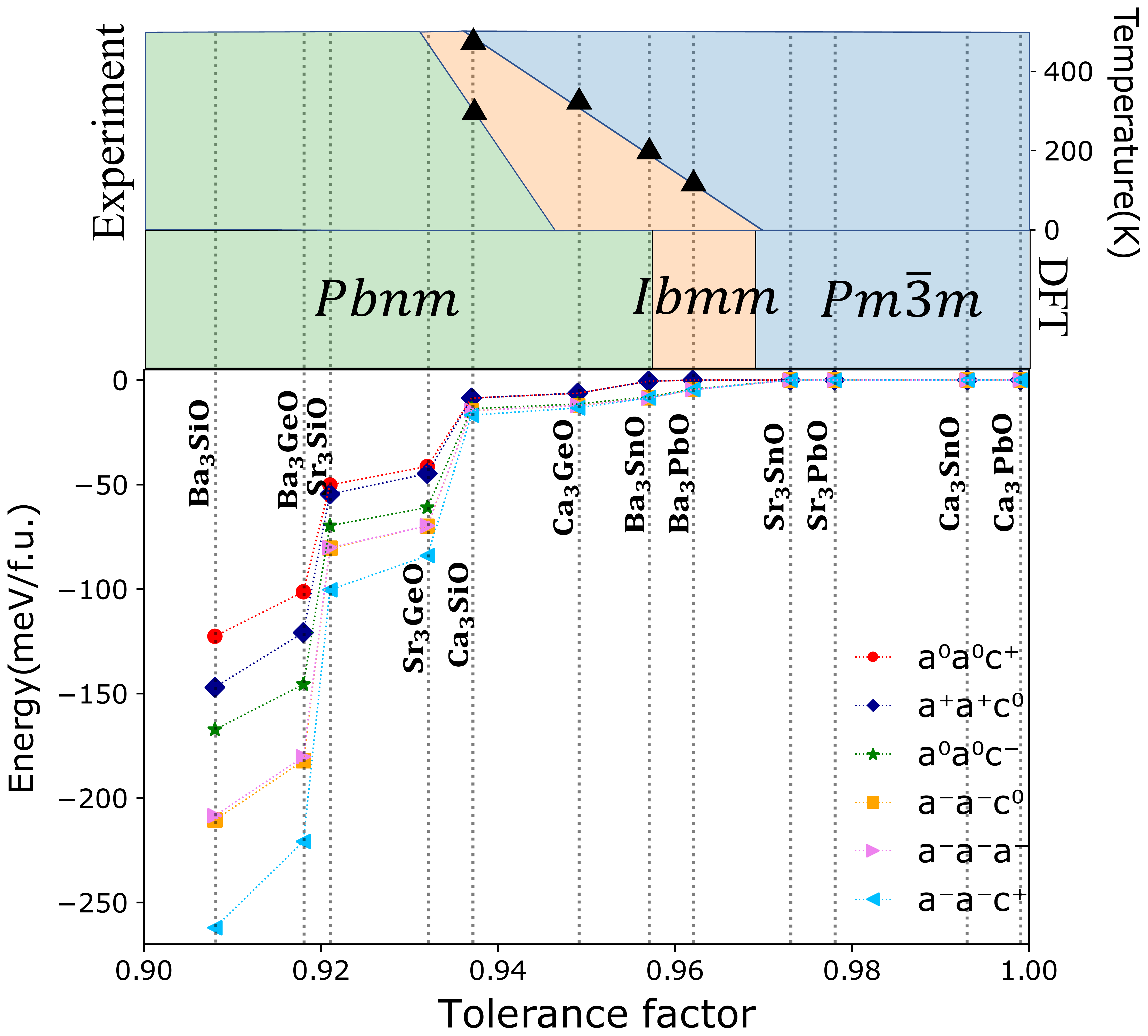}}
\caption{Phase diagram of octahedral rotations. The colored areas represent $Pbnm$, $Ibmm$ or $Pm\bar3m$ phases, predicted by experimental results and DFT calculations. The phase transition temperatures are denoted by black triangles. Below them are the details of DFT calculations: The energies of tilted structures are given relative to energies of cubic phase $Pm\bar3m$. The experimental data are reproduced from \cite{nuss2015tilting}.} \label{fig:Phases}
\end{figure}

Nuss et al. conducted single crystal X-ray diffraction experiments for X$_3$TtO series \cite{nuss2015tilting}. Within the temperature range 50-500 K, Ca$_3$SiO, Ca$_3$GeO, Ba$_3$SnO and Ba$_3$PbO undergo a phase transition from cubic ($Pm\bar3m$, $\#221$) to orthorhombic ($Ibmm$, $\#74$); Ca$_3$SiO undergoes an additional phase transition to the GdFeO$_3$-type orthorhombic structure ($Pbnm$, $\#62$); all other AP oxides retain either cubic or $GdFeO_3$-type structures in this temperature range. $Ibmm$ and $Pbnm$ phases result from octahedral rotation $a^-a^-c^0$ and $a^-a^-c^+$, respectively. These experimental observations of Ref.~\cite{nuss2015tilting} are summarized in the top row of Fig. \ref{fig:Phases}, along with the phase transition temperatures they measured. Notably, the intermediate phase between the cubic and the orthorhombic structure that is commonly observed in oxide perovskites, i.e. the tetragonal phase with $a^0a^0c^-$ rotation pattern, was not observed in \cite{nuss2015tilting}. This is peculiar because in the perovskites with octahedral rotational instabilities, a direct phase transition from $Pm\bar3m$ to $Ibmm$ phase is rare, and often an intermediate tetragonal phase is observed \cite{howard2000structural, mountstevens2005order, ma2008perovskite, carpenter2006structural}. The Phase transition temperatures in Fig. \ref{fig:Phases} highlight the significance of size effect in determining the stabilities of different structures. However, due to the limitations of the temperature range in the experiment of \cite{nuss2015tilting}, it is likely that not all the possible 0 temperature polymorphs are captured, thus that the structure field map is only a rough estimate. In order to confirm the experimental observations and predict the low temperature ground state structures, we apply DFT calculations and elucidate how the tolerance factor affects the ground-state structures of AP oxides with tetrel ions. 

\begin{figure}[b]
\includegraphics[height = 15.0cm]{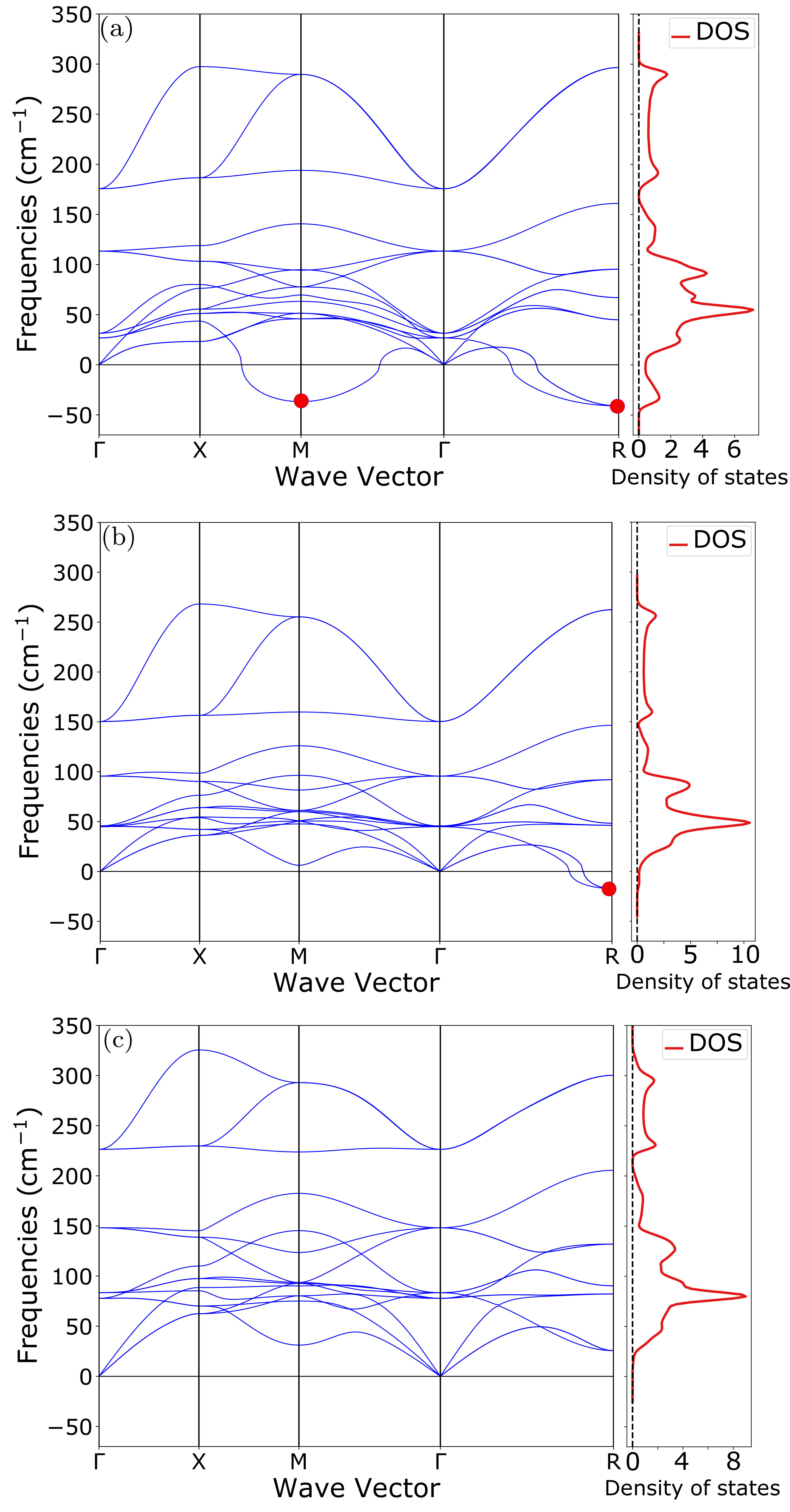}
\caption{The phonon dispersion relations (left) and total phonon density of states (right) for cubic (a) Ba$_3$GeO, (b) Ba$_3$PbO, (c) Sr$_3$SnO. Unstable modes at high symmetry points M and R, which are discussed in the text, are denoted by red circles.} \label{fig:Phonon}
\end{figure}

We start by examining the phonon dispersions in cubic phases of each compound. The dispersions of many compounds are similar, and in Fig. \ref{fig:Phonon}, we present three examples. The `+' and `-' rotations correspond to $M^+_2$ and $R^-_5$ phonon modes, respectively. The most unstable modes were found in compounds with the lowest tolerance factor, such as silicon or germanium APs. Fig. \ref{fig:Phonon}(a) shows the phonon dispersions of Ba$_3$GeO with tolerance factor 0.918. The frequencies of both $M^+_2$ and $R^-_5$ instabilities increase with increasing tolerance factor (Table~\ref{tab:Phonofreq}). The $M^+_2$ mode becomes positive first, e.g. in Ba$_3$PbO with tolerance factor 0.962 (Fig. \ref{fig:Phonon}(b)), and then $R^-_5$ becomes stable too, e.g. in Sr$_3$SnO with tolerance factor 0.973 (Fig. \ref{fig:Phonon}(c)). Table \ref{tab:Phonofreq} summarizes the lowest frequencies of $M^+_2$, $R^-_5$ modes of all AP oxides we studied.

\begin{figure*}[bht]

\includegraphics[height = 7cm]{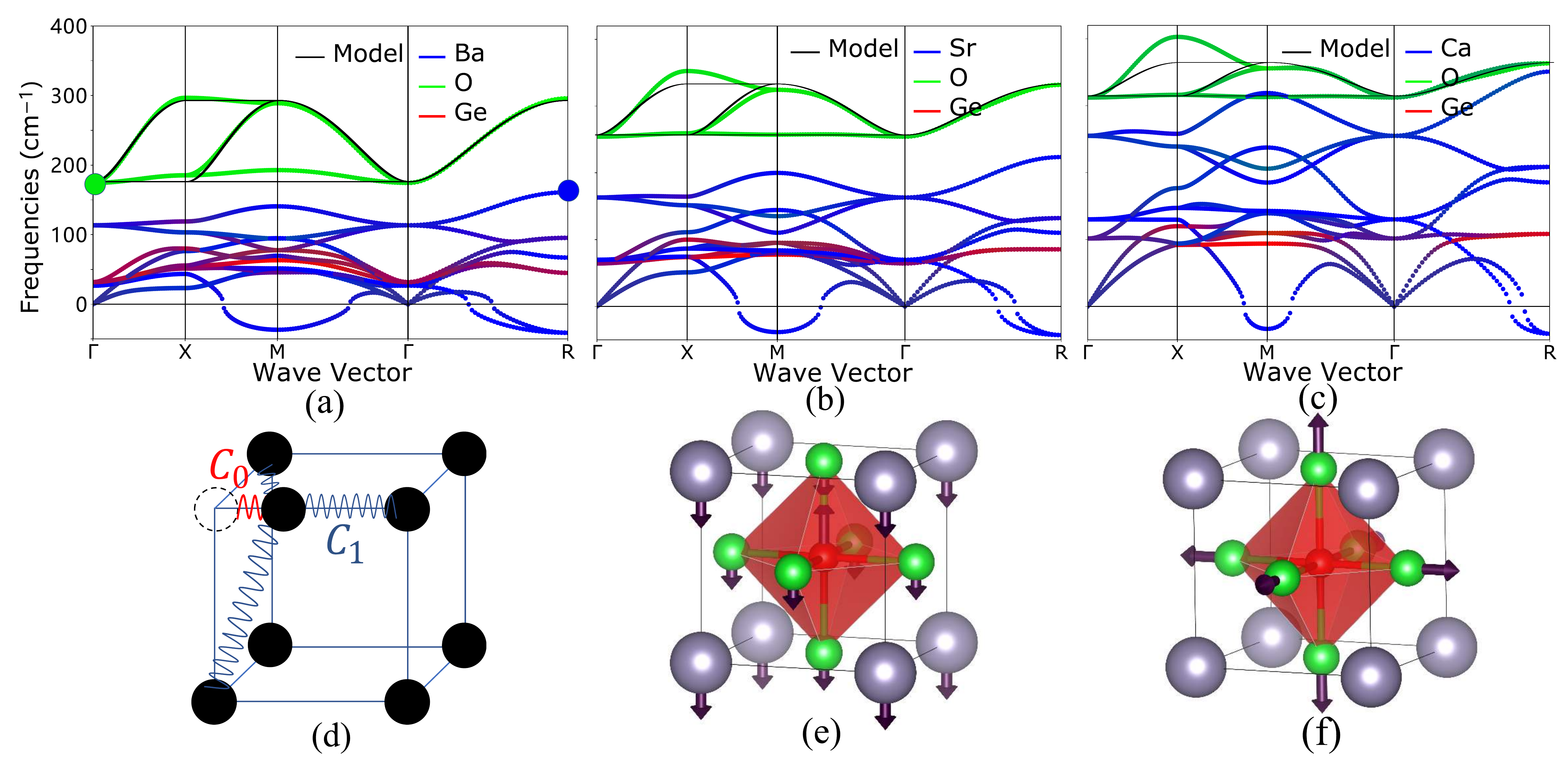}
\caption{\label{fig:Proj_Phonon} (a)-(c) the atom-resolved phonon band structures for germanium APs; The oxygen band predicted by the model in (d) is overlaid; The green and blue dots label the polar mode and the breathing mode, respectively. (d) Simplified model for oxygen vibration sampled on a cubic lattice. $C_1$ represents the force constant between two adjacent oxygen atoms, and $C_0$ represents the restoring force constant in order to avoid linearity near the $\Gamma$ point. (e) Schematic for polar modes in proper ferroelectric perovskites. (f) Schematic for the breathing mode.}
\end{figure*}

In addition to indicating structural instabilities, the projected phonon band structures shown in Fig.~\ref{fig:Proj_Phonon} display other interesting properties as well. The  oxygen bands always have the highest frequencies, which can be explained by the low mass of oxygen ions compared to the other atoms. However, unlike oxide perovskites where the oxygen-dominated phonons often have frequencies as high as $\sim 800$~cm$^{-1}$, the highest frequency phonons are much softer, attesting to the lower force constants in APs in general. In barium and strontium APs, the oxygen bands are isolated from other bands by a gap as large as $\sim 50$~cm$^{-1}$, as shown in Fig. \ref{fig:Proj_Phonon}(a) and (b). The Slater mode, which consists of the body-center atom vibrating against the face-center ones (Fig. \ref{fig:Proj_Phonon}(e)), has the highest frequency in APs, even though it is often the softest polar mode in perovskites. This is likely a consequence of smaller mass of oxygen atoms. Fig. \ref{fig:Proj_Phonon}(c) shows that in calcium compounds, oxygen bands cross calcium bands near the M point, yielding a potential nontrivial topology and a phonon nodal line.

The shape of oxygen bands indicates that the motion of oxygen atoms has only a small impact on the other two types of atoms. To show this, we build a toy model of lattice vibrations for a single oxygen atom in a simple cubic lattice, which comprises only two free parameters reflecting the nearest-neighbor force constant ($C_1$) and the self-force-constant ($C_0$). (The mass of the oxygen ion can be considered as just a rescaling of these parameters.) This model, shown in Fig.~\ref{fig:Proj_Phonon}(d), reproduces the main features of oxygen bands with surprising quality, including the flat band along $\Gamma$-$X$-$M$, underlining that the oxygen atom vibrations are decoupled from the rest of the crystal to a large extent. 

In passing, we note that the breathing mode (Fig. \ref{fig:Proj_Phonon}(f)) is the hardest mode controlled by alkaline metal displacements, but it is still dramatically softer than the oxygen breathing mode in perovskites. 

In the lower panel of Fig. \ref{fig:Phases}, we show our DFT results for the energies of APs in different phases ($a^-a^-c^+$, $a^0a^0c^+$, $a^+a^+c^0$, $a^0a^0c^-$, $a^-a^-c^0$ and $a^-a^-a^-$) relative to the energy of the cubic phase. The structures are obtained by fully relaxing the atomic positions as well as lattice parameters in the given space groups. For all compounds with the tolerance factor smaller than 0.95, the energies of $a^-a^-c^+$ structures are lower than any other structure. For $0.95 < t_a <0.97$, the $c^+$ rotations tend to vanish and the $a^-a^-c^0$ tilting pattern has the lowest energy. Similarly, for those with $t_a > 0.97$, all the octahedral rotations vanish, and the cubic structures are the ground state structures. The ground state phases DFT predicts are in agreement with the experimental results to a large extent. The $Pbnm$ phase we predict for Ca$_3$GeO and Ba$_3$SnO is not observed in the experiments in Ref.~\cite{nuss2015tilting}, likely because the experiments did not go below 50~K. The DFT energy difference in these compounds between the $Pbnm$ and $Ibmm$ phases is less than this temperature scale divided by $k_B$, indicating that a temperature of 50~K could possibly suppress this phase. 
Similarly, while no phase other than the orthorhombic $Pbnm$ is reported for compounds such as Sr$_3$GeO and Sr$_3$SiO, the DFT energy scales suggest transitions above 500 K but below the melting temperature of these systems. The ground-state structures obtained from DFT energy calculations and in experiments are also summarized in Table \ref{tab:stability}. 


\begin{table}[bht]
\begin{tabular}{c c c c c c}
\hline
Compounds  & $t_a$ &   $M^+_2$& \quad $R^-_5$ &\quad DFT / Experiment \\ \hline

  Sr$_3$SiO& 0.921 &  $40\, i$& \quad $44\, i$ &\quad$Pbnm$ / $Pbnm$ \\
  
  Sr$_3$GeO& 0.932 & $39\, i$& \quad $43\, i$ &\quad $Pbnm$ / $Pbnm$ \\
  
  Ba$_3$SiO& 0.908 &$38\, i$ &\quad $42\, i$  &\quad $Pbnm$ / $Pbnm$ \\
  
  Ba$_3$GeO& 0.918 &$37\, i$ &\quad $41\, i$  &\quad $Pbnm$ / $Pbnm$ \\
  
  Ca$_3$SiO& 0.937 &$34\, i$ &\quad $41\, i$  &\quad $Pbnm$ / $Pbnm$\\
  
  Ca$_3$GeO& 0.949 &$34\, i$ &\quad $41\, i$  &\quad $Pbnm$ / $Ibmm$\\
  
  Ba$_3$SnO& 0.957 &$9\,  i$ &\quad $19\, i$  &\quad $Ibmm$ / $Ibmm$\\
  
  Ba$_3$PbO& 0.962 &$6   $ &\quad $17\, i$  &\quad $Ibmm$ / $Ibmm$\\
  
  Sr$_3$SnO& 0.973 &$31$  &\quad $26$  &\quad $Pm\bar3m$ / $Pm\bar3m$\\
  
  Sr$_3$PbO& 0.978 &$32$& \quad $26$  &\quad $Pm\bar3m$ / $Pm\bar3m$\\
  
  Ca$_3$PbO& 0.999 &$65$ &\quad $68$  &\quad $Pm\bar3m$ / $Pm\bar3m$\\
  
  Ca$_3$SnO& 0.993 &$72$ &\quad $69$  &\quad $Pm\bar3m$ / $Pm\bar3m$\\
\end{tabular}
\caption{The tolerance factor $t_a$, frequencies of the least stable $M^+_2$, $R^-_5$ modes, and ground-state structures obtained from both DFT calculations and experiments for X$_3$TtO. $i$ is the imaginary unit. The unit of frequencies is $cm^{-1}$.} \label{tab:Phonofreq}
\end{table}

\begin{figure*}[bht]
\includegraphics[height = 6cm]{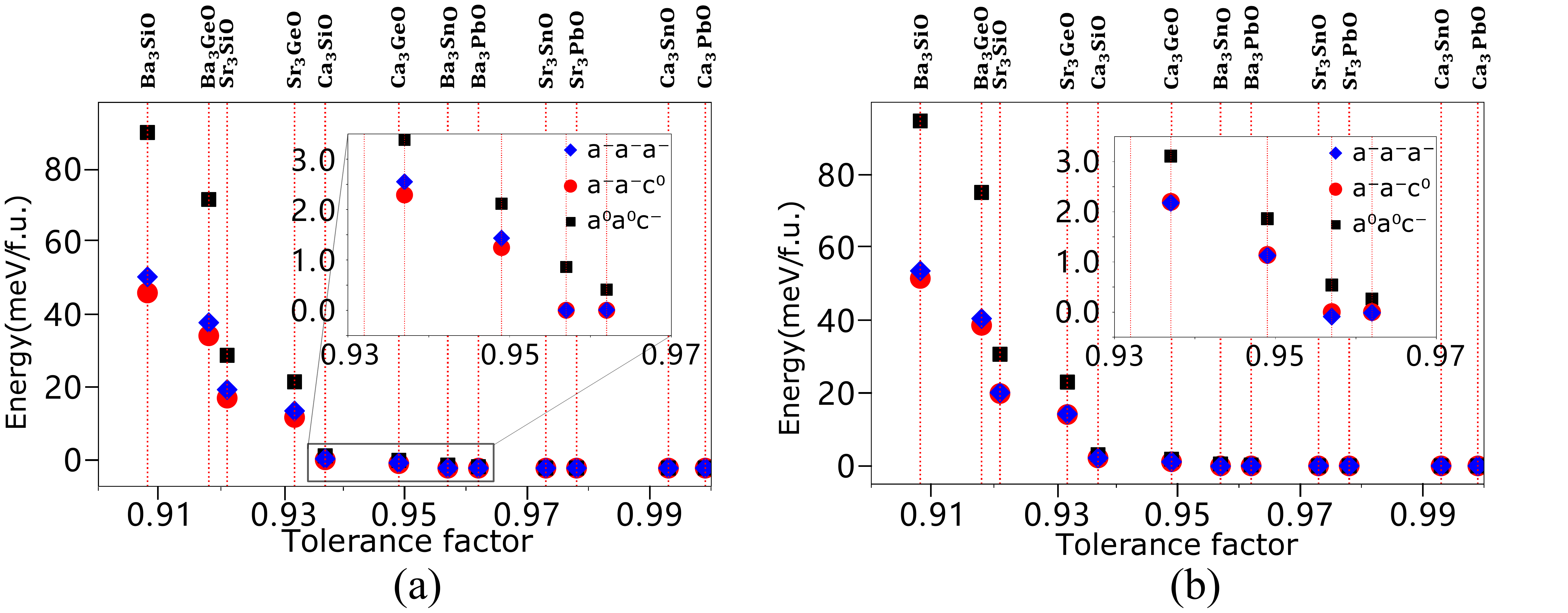}
\caption{\label{fig:three_phases}Diagrams showing energies for APs with only out-of-phase octahedral rotations with (a) fixed lattice parameters (b) relaxed lattice parameters. The energies are relative to $a^-a^-c^+$ systems. The insets are the zoom of the boxed area that displays the energies of Ca$_3$SiO, Ca$_3$GeO, Ba$_3$SnO, and Ba$_3$PbO. }
\end{figure*}

It is worth noting that just like most perovskite oxides, which don't have a cubic structure at low T \cite{woodward1997octahedralII, mulder2013turning}, most of the APs with tetrel and oxide ions we considered have octahedral rotations and a non-cubic symmetry. The critical tolerance factor for the emergence of rotations is $\sim 0.97$. This indicates that either cubic AP oxides are more stable against undersized A-site anions, or the $Tt^{-4}$ radii we employed are not accurate, as discussed earlier. While it is not possible to reach a conclusion in the absence of more data to obtain better $Tt^{-4}$ radii, we note that earlier studies on AP nitrides and fluorides display a similar trend \cite{mochizuki2020theoretical, fujii2021alkali}.

An interesting trend in the energies of different phases is that especially for materials with $t_a>0.93$, the single, double, and triple out-of-phase tilt systems (i.e. $a^0a^0c^-$, $a^-a^-c^0$, and $a^-a^-a^-$ phases) are very close in energy. In Fig.~\ref{fig:three_phases}, we show the energies of these three phases with respect to the lowest energy phase, which makes it evident that the energy difference between $a^-a^-c^0$ and $a^-a^-a^-$ phases are well less than 1~meV/f.u. (formula unit) even when strain degrees of freedom are taken into account by relaxing the lattice vectors as well. The fact that the $a^0a^0c^-$ phase is also within 1 meV/f.u. might explain why this tetragonal $I4/mcm$ phase is not experimentally observed in these compounds. 

\begin{figure}[bht]
\includegraphics[width=0.98\linewidth]{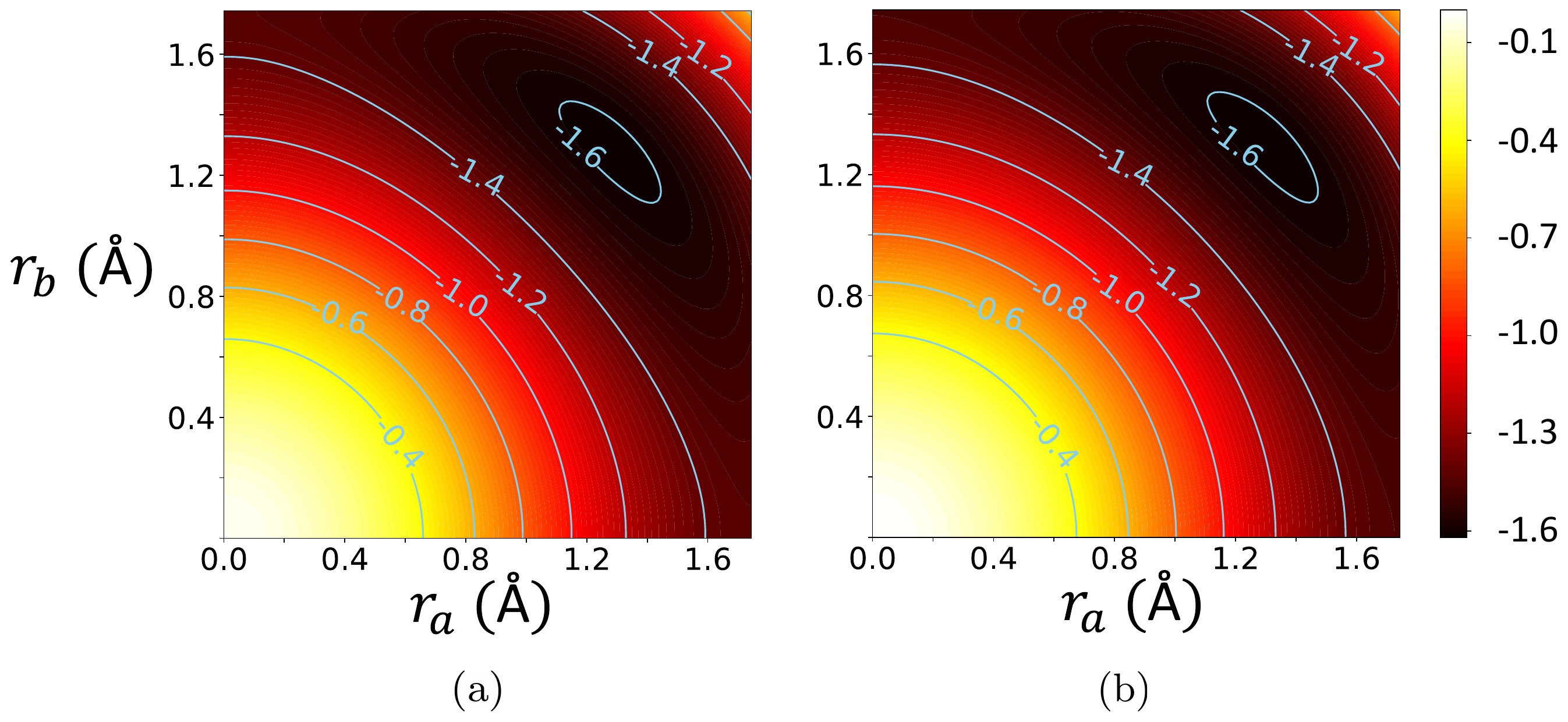}
\caption{\label{fig:fit_data}Energy landscapes for SrSnO$_3$ in terms of x-axis and y-axis out-of-phase rotation, where the energies are (a) directly obtained from DFT calculation; (b) calculated by fitting LFE up to the fourth order terms. The energies are given relative to the reference phase $a^0a^0a^0$, in meV/f.u.}
\end{figure}

To explain this trend, which is not observed in perovskite oxides, we examine the Landau free energy (LFE) of these systems. We consider only the $R_5^-$ mode, and represent its three components (rotations around three crystallographic axes) as $\bm{r}=(r_a, r_b, r_c)$. The energy with respect to the cubic phase has the form 
\begin{equation}
E = \alpha |\bm{r}|^2 + \beta |\bm{r}|^4 + \gamma(r^2_ar^2_b+r^2_br^2_c+r^2_cr^2_a), \label{eq::LFE_general}
\end{equation}
where $\alpha$ and $\beta$ are materials-specific coefficients that can be obtained from DFT. Minimizing the free energy with respect to $r$ for the $a^0a^0c^-$ system ($r_c = r$ and $r_a = r_b = 0$) gives 
\begin{equation}
r_{a^0a^0c^-} = \sqrt{-\frac{\alpha}{2\beta}},
\end{equation}
and 
\begin{equation}
E_{a^0a^0c^-} = -\frac{\alpha^2}{4\beta}.  \label{eq::LFE_a0a0c-}
\end{equation}
Similarly, the mode amplitudes and energies for the energy minima corresponding to $a^-a^-c^0$ and $a^-a^-a^-$ systems are
\begin{equation}
\begin{aligned}
r_{a^-a^-c^0} &= \sqrt{-\frac{\alpha}{4\beta+\gamma}};\\
E_{a^-a^-c^0} &= -\frac{\alpha_r^2}{4\beta + \gamma};\\
r_{a^-a^-a^-} &= \sqrt{-\frac{\alpha}{6\beta+2\gamma}};\\
E_{a^-a^-a^-} &= -\frac{\alpha_r^2}{4\beta + \frac{4}{3}\gamma}. \label{eq::LFE_a-a-c0_a-a-a-}
\end{aligned}
\end{equation}
The only term in the free energy that differentiates between different directions of the order parameter $\bm{r}$ is $\gamma$. In particular, if the dimensionless parameter $\gamma/\beta$ is small, then the energies of the three out-of-phase tilt systems become degenerate. To show that this is indeed the case in the APs under question, we calculated the free energy coefficients of these compounds from DFT. In particular, we performed single-shot self-consistent DFT calculations for different crystal structures where the octahedra are rotated around different axes by different amounts. While it is not necessary to perform a large number of calculations for this, we considered a total of 64 structures on a equally-space 2D grid of rotation amplitudes around $a$ and $b$ axes. For comparison, we also perform similar calculations for a number of perovskite oxides. As shown in Fig.~\ref{fig:fit_data} for SrSnO$_3$, the 4th order polynomials with the fitted coefficients order are sufficient to reproduce the DFT energies. In Table \ref{tab:coefficients}, we tabulate the coefficients of APs Ca$_3$SiO, Ca$_3$GeO, Ba$_3$SnO, and Ba$_3$PbO, as well as representative perovskites, SrSnO$_3$, CaSnO$_3$, CaGeO$_3$, and BaPbO$_3$. (The $Ibmm$ to $I4/mcm$ phase transition in SrSnO$_3$ is well established \cite{mountstevens2005order}.)

\begin{table*}[bht]
\renewcommand\arraystretch{1.2}
\caption{Coefficients for LFE obtained by fitting DFT energies. The coefficients are calculated so that the $\Delta E$ gives energy shifts from cubic phases per f.u.} \label{tab:coefficients}
\begin{tabular}{c c c c c c c c c}
\hline

Compounds & Ca$_3$SiO & Ca$_3$GeO & Ba$_3$SnO & Ba$_3$PbO & \quad \quad SrSnO$_3$ & CaSnO$_3$ &CaGeO$_3$ & BaPbO$_3$\\ \hline
  structure & $Pbnm$ & $Pbnm$ & $Ibmm$ & $Ibmm$ & \quad \quad $Pbnm$ & $Pbnm$ & $Pbnm$ & $Ibmm$ \\  
  $\alpha (meV/\AA^2)$ & $-19.4$ & $-17.9$ & $-11.4$ & $-8.4$ & \quad \quad $-118.8$ & $-191.5$ & $-171.0$ & $-82.9$ \\
  $\beta (meV/\AA^4)$& $7.6$ & $7.5$ & $4.5$ &$4.4$&  \quad \quad $18.8$ & $18.5$ & $32.3$ & $13.9$ \\
  $\gamma (meV/\AA^4)$& $-0.8$ & $-0.8$ & $-1.0$ & $-1.0$ & \quad \quad $-7.1$ & $-14.8$ & $-19.1$ & $-6.0$ \\
  $|\gamma/\beta|$\% & $10.5$ & $10.7$ & $22.2$ & $22.7$ &\quad \quad $37.8$ & $80.0$ & $59.1$ & $43.2$
\end{tabular}

\end{table*}

In both perovskite and AP compounds, the negative sign of $\gamma$ ensures $E_{a^0a^0c-} > E_{a^-a^-c^0} > E_{a^-a^-a^-}$, consistent with the results of structural relaxations in Fig. \ref{fig:Phases} and Fig. \ref{fig:three_phases}. However, in APs,  $|\gamma/\beta|$ is smaller compared to perovskites, especially in Ca$_3$SiO and Ca$_3$GeO. This explains the near-degeneracy of different tilt systems in APs. 
The absence of the single-tilt ($I4/mcm$) structure is likely related to the small $|\gamma/\beta|$ as well: This number not only determines the $a^0a^0c^-$ and $a^-a^-c^0$ structures' energies, but it also determines the energy barrier between these two phases. Hence, a small $|\gamma/\beta|$ should reduce any temperature hysteresis between these two phases, which might make the single-tilt structure constrained only at a much narrower or disappearing temperature range. 

In passing, we note that the magnitude of the LFE parameters depend on the normalization of the order parameters. We normalized our mode amplitudes in the $2\times 2 \times 2$ supercell by using the definition:
\begin{equation}
r_i = \sqrt{\sum_{j}u_{j,i}^2},
\end{equation}
where $i = \{a,b,c\}$, $j$ indexes the atoms in the supercell.

\subsection{Hybrid Improper Ferroelectricity in Antiperovskites}

There is no combination of octahedral rotations that break the inversion symmetry and leads to ferroelectricity in perovskites. However, when combined by some other symmetry breaking effects, octahedral rotations or their combinations may become polar \cite{benedek2012polar, li_suppressing_2020}. One way of realizing ferroelectricity in perovskites through octahedral rotations is to consider cation-ordered systems \cite{bousquet2008improper, rondinelli2012octahedral, swamynadhan_design_2023}, also referred to as double perovskites. Another way is to consider Ruddlesden-Popper phases, which can be considered as layered perovskites with ordered stacking faults. The n'th Ruddlesden-Popper structure consists of n perovskite layers followed by an additional $A-X$ layer, and has the general chemical formula A$_{n+1}$B$_n$X$_{3n+1}$ \cite{ruddlesden_compound_1958}. In both of these families, pairs of octahedral rotations give rise to a spontaneous and switchable polarization through the hybrid-improper mechanism \cite{benedek2011hybrid}. 
While similar phenomenological mechanisms where the combination of two nonpolar modes give rise to ferroelectricity are also relevant to other classes of materials, e.g.  HfO$_2$ \cite{jung2025triggeredferroelectricityhfo2hybrid, jung2025electricpolarizationnonpolarphonons} and Dion-Jacobson phases \cite{cascos_tuning_2020}, the ubiquitous nature of the octahedral rotations in perovskites make them a particularly fertile family to design and discover new ferroelectrics \cite{zhang_review_2022, benedek_hybrid_2022}. In this section, we show that the mechanism of hybrid improper ferroelectricity is also active in APs, and it can be used to obtain inversion-breaking variants of these materials which in their bulk form don't exhibit any ferroelectricity. While there is one study which predicted the possibility of hybrid improper ferroelectricity in an antiperovskite heterostructure \cite{garcia2019hybrid}, there is no systematic study on ferroelectricity in the tetrel oxide APs that elucidate trends to the best of our knowledge, and in this section, we aim to ameliorate this. 

\subsubsection{Anti-Ruddlesden-Popper phases}

There are many works on Anti-Ruddlesden-Popper (ARP) nitrides, their structure, and chemical properties spanning decades \cite{rohr1996crystal, hadenfeldt1988darstellung, hadenfeldt1991darstellung, gaebler2007first, wied2011crystal}. Among these, those on ferroelectic Ba$_4$(Sb, As)$_2$O and antiferroelectric Ba$_4$(Bi, P)$_2$O and Sr$_4$(Bi, Sb, P)$_2$O are of particular interest \cite{markov2021ferroelectricity}. However, tetrel elements and their oxides we consider here do not commonly form ARP structures. This is mainly due to the nominal charges of these ions. For example, the $n=1$ ARP Ba$_4$Sn$_2$O would require Sn to have a charge of $-3$, which is not impossible, but it is extremely rare and reported only in compounds with Zintl bonding \cite{Papoian2000}. The $n=2$ Ba$_7$Sn$_3$O$_2$ would be even less likely to form, since it would necessitate a non-integer charge for Sn. To confirm these predictions, we built hypothetical X$_{3n+1}$Tt$_{n+1}$O$_n$ compounds with $n = 1, 2, 3$ using the tetrel and alkaline earth elements, performed structural relaxations using DFT, and built convex hulls \cite{peterson2021materials} to predict thermodynamic stability from first principles. The energies of these ARP phases with respect to the convex hull are shown in the supplementary information \cite{supplement}. Without exception, in all cases we considered, none of the ARP phases were thermodynamically stable, and they decompose into different chemical phases. As a result, we do not consider ARP phases further.

\subsubsection{Double Antiperovskites (DAPs)}

\begin{figure*}[bht]
\includegraphics[width=0.9\linewidth]{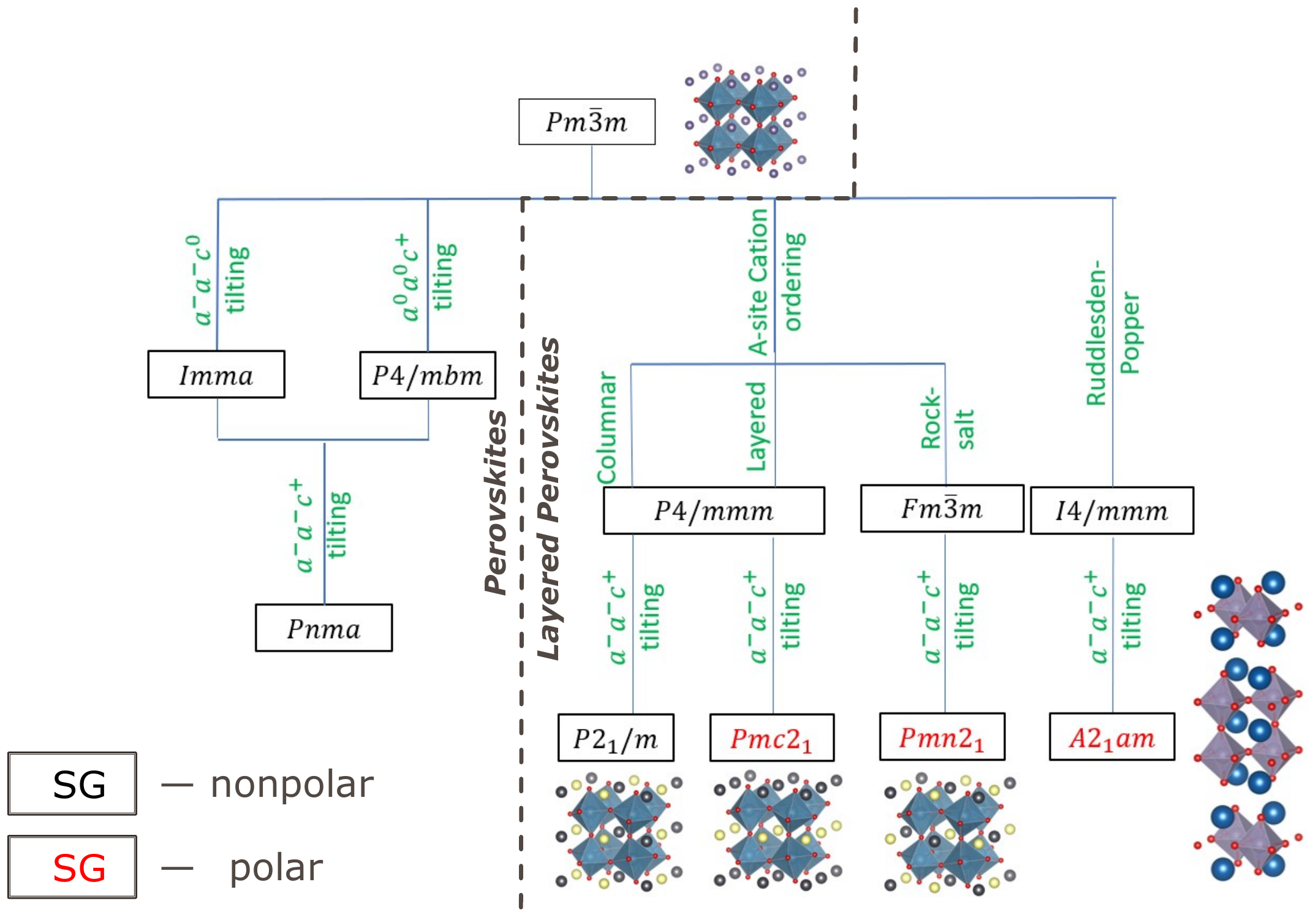}
\caption{\label{fig:tree} Tree diagram depicting group-subgroup relationship between different structure and cation ordered phases of perovskite and perovskite-like structures. Polar space groups relevant to ferroelectricity are shown in red. }
\end{figure*}

\begin{figure}[bht]
\includegraphics[height = 3.1cm]{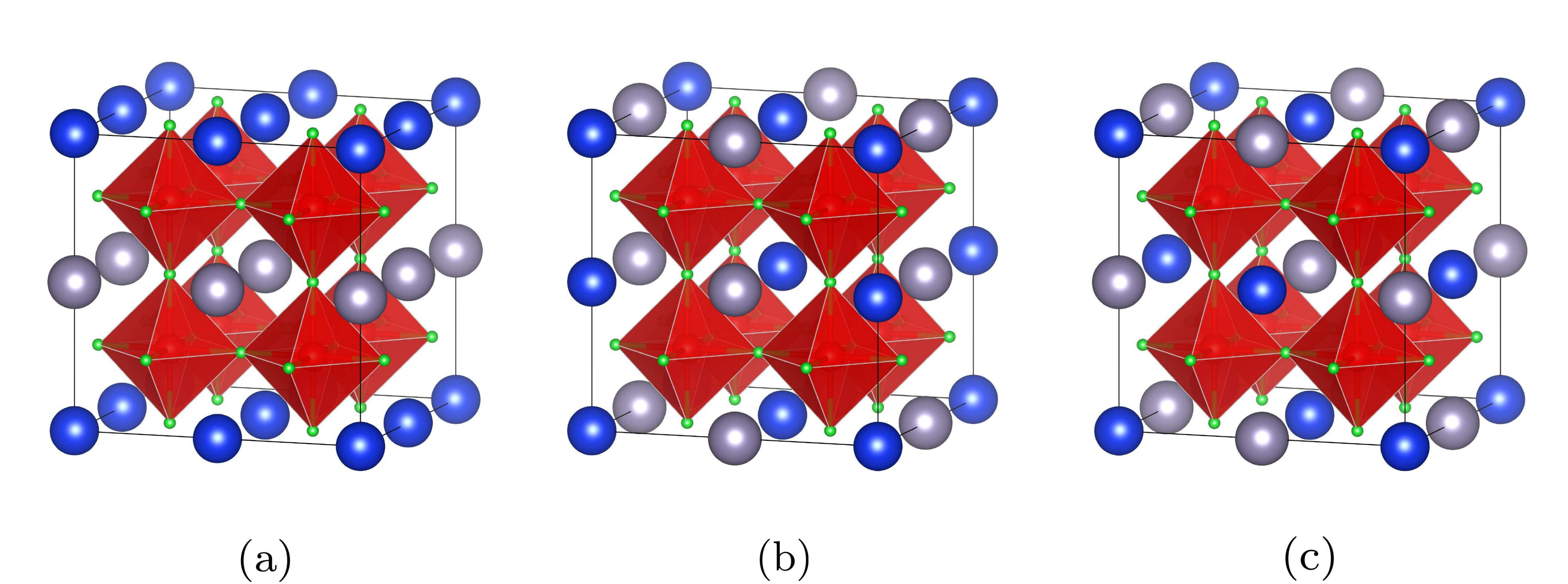}
\caption{\label{fig:patterns}Schematic showing A-site ordering of double APs (or perovskites) with (a) layered ordering, (b) columnar ordering, and (c) rock-salt ordering. Blue and grey spheres represent the two inequivalent A site ions. }
\end{figure}

In Fig.~\ref{fig:tree}, we summarize the relevant structures with octahedral rotations in perovskites (or equivalently APs) and double (anti-)perovskites, along with their corresponding space groups. The three common types of cation ordering (columnar, layered, and rock-salt) are shown in Fig.~\ref{fig:patterns}. 
In double perovskites, not all cation orders are conducive to hybrid-improper ferroelectricity, and this phenomenon is mostly studied in the layered-ordered phases \cite{rondinelli2012octahedral, gayathri_predictive_2024}. While this ordering pattern is the most common one in A-site ordered double-perovskites \cite{king2010cation}, the question whether it is preferable in double APs is not yet answered conclusively. 

There isn't extensive work on ordered DAP's, but, for example, a rock-salt-ordered nitrohalide DAPs was successfully synthesized \cite{mi2022rock}, and there is no fundamental reason for cation ordering to be rare in DAPs.
While theory studies that compare different cation orders in DAPs are also rare, Ref.~\cite{Han2021} theoretically found that  layered-ordering is preferable in nitride DAPs \cite{Han2021}. In this section, to address the question of cation ordering in materials of interest, we consider different cation ordering possibilities in double perovskites, and apply the theoretical machinery developed for double perovskites to double APs to predict if they stabilize hybrid-improper ferroelectricity.

Analogous to A-site ordered AA'B$_2$X$_6$ double perovskites \cite{king2010cation}, we consider DAPs with A-site orderings with the chemical formula X$_6$AA'O$_2$, where the A-site anions can order into layered, columnar, or rock-salt patterns  (Fig. \ref{fig:patterns}). 
Before studying polarization in these compounds, we first determine which ordering pattern has the lowest energy. In Table~\ref{tab:stability}, we report the energies of different ordering patterns as predicted by DFT calculations performed for each of these structures. For all of the tetrel DAP oxides  considered, the layered ordering pattern leads to the lowest energy. This can be explained in the same way that the prevalence of A-site layered ordering is explained in perovskites \cite{king2010cation}, following mostly ion-size and electrostatic arguments. 

Comparing the energy difference between cation-ordered phases with the average energies of individual single AP phases shows that DAPs are thermodynamically unstable, and under thermodynamic equilibrium, they would phase separate to single APs. However, this energy scale is as low as a few meV/f.u. for certain DAPs, which suggests that DAPs can be possibly stabilized through different growth and processing procedures in bulk (as is often the case in double perovskites) or by layer-by-layer growth \cite{hossain2018overview, todate2007magnetic, mclaughlin2006magnetic, wakeshima1999magnetic}.

\begin{table}[bht]
\renewcommand\arraystretch{1.2}
\caption{Relative energies of cation orderings, and the averaged energy of the corresponding AP composites per formula unit, given in meV. Ten DAPs(without octahedral rotation) are included. Energy for layered ordering is set to be the references so that $\Delta E_x = E_x - E_{layered}$. $RS =$ rock-salt and $C =$ columnar.} \label{tab:stability}
\begin{tabular}{c| c| c| c}
\hline
Compound & $\Delta E_{RS}$ & $\Delta E_{C}$ &  $\Delta [(E_{X_3TtO} + E_{X^{'}_3TtO})/2]$  \\ \hline
  Ba$_6$SiGeO$_2$ & +1.1 & +0.4 & -3.0    \\
  Ba$_6$SiPbO$_2$ & +42.7 & +13.3 & -16.0 \\
  Ba$_6$SiSnO$_2$ & +25.4 & +7.6 & -11.0 \\
  Ba$_6$SnPbO$_2$ & +2.2 & +0.8 & -4.8 \\
  Ca$_6$SiPbO$_2$ & +64.7 & +24.3 & -41.9 \\
  Ca$_6$SnPbO$_2$ & +2.6 & +1.0 & -11.1 \\
  Sr$_6$SiGeO$_2$ & +1.2 & +0.5 & -11.1 \\
  Sr$_6$SiPbO$_2$ & +50.7 & +18.2 & -33.7 \\
  Sr$_6$SnPbO$_2$ & +2.7 & +1.1 & -11.3 \\
  Sr$_6$SiSnO$_2$ & +32.2 & +11.2 & -22.6 \\
\end{tabular}
\end{table}

Finding that the layered ordering is energetically preferred, we now focus on tetrel DAPs with the A-site layered ordering only, and consider the possibility of ferroelectricity induced by octahedral rotations. In the remainder of this paper, DAP refers to structures with this specific cation order. Interestingly, in all DAPs we consider, the $a^+b^-c^-$ rotation pattern, as opposed to the $a^-a^-c^+$ one, is energetically favorable. (Here, the layering direction is selected as $c$.) This rotation pattern does not lead to a polar group, and as a result, the as grown bulk DAPs may not be conducive to ferroelectricity. However, the typical energy difference is less than 10~meV/f.u., which suggests that field cooling using an in-plane electric field may favor the stabilization of a metastable ferroelectric phase. In the next subsection, we will show that moderate compressive strain drives the barium DAPs to the polar $Pmc2_1$ phase with $a^-a^-c^+$ rotation pattern, which is another promising avenue to observe ferroelectricity in DAPs. 

\begin{figure}[bht]
\begin{center}
\includegraphics[height = 4.5cm]{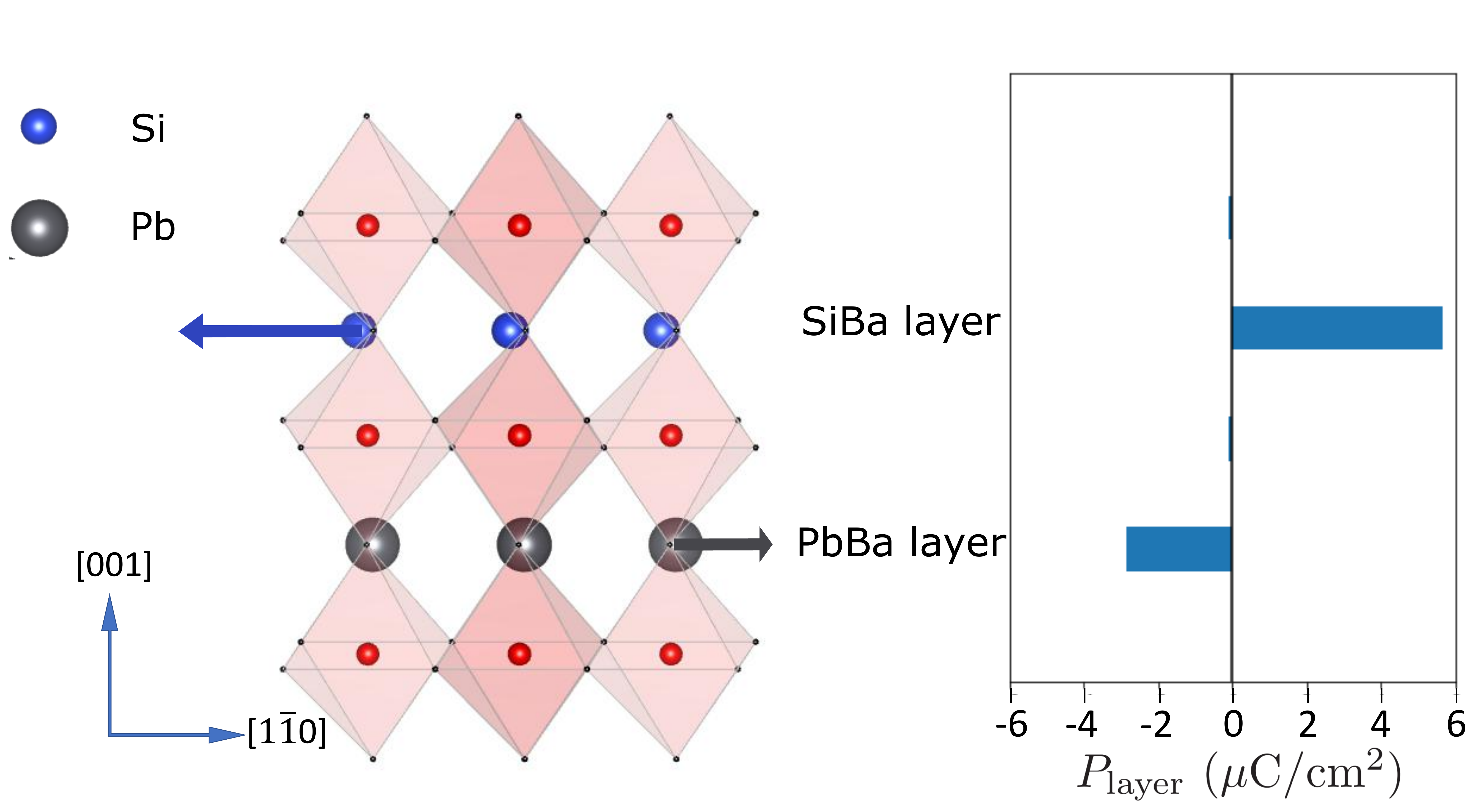}
\caption{\label{fig:LayPol} Mechanism of improper ferroelectrics in DAPs, in particular, Ba$_6$SiPbO$_2$ as an example. Left: the displacements of A-site anions. The lengths of the arrows reflect the amplitudes of displacements. The shift of charge center of OBa$_2$ layers are non-zero, but smaller by more than an order of magnitude than that of the PbBa layers, and hence not shown. Right: layer-resolved polarization, including that of the BaO$_2$ layers.}
\end{center}
\end{figure}

To study the correlation between the tolerance factor and the net polarization in the polar hybrid-improper ferroelectric phase ($Pmc2_1$), we performed structural relaxations for 18 DAP oxides in this phase. Fig. \ref{fig:LayPol} demonstrates the microscopic mechanism of ferroelectric in Ba$_6$SiPbO$_2$, which is anologous to similar diagrams in double perovskites \cite{mulder2013turning}. The main contribution to net polarization is the unequal anti-polar displacements along $[1\bar10]$ between A-sites and A'-sites. The off-centering displacements of oxygen atoms are minimal. Almost all strontium and barium based DAPs we consider inherit robust octahedral rotations from single APs, resulting in non-zero net polarization in this phase. However, among calcium DAPs, only Ca$_6$SiGeO$_2$, whose average tolerance factor is the lowest among all the calcium APs, has non-zero spontaneous polarization.

\begin{figure*}[bht]
\includegraphics[height = 6cm]{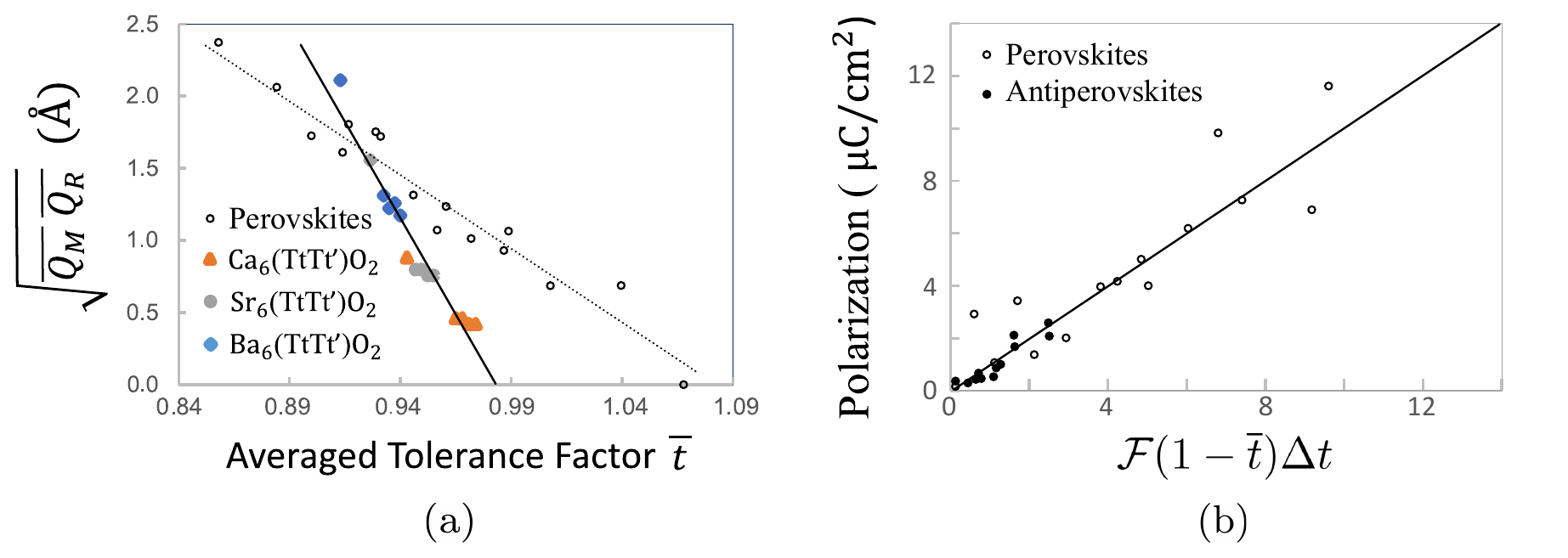}
\caption{\label{fig:Pol_Mode_vs_tol} (a) The relationship between the tolerance factor and octahedral rotation amplitudes in double perovskites (black open circles) and DAPs (filled markers). The dashed and solid lines represent the linear functions that fit data of perovskites and APs, respectively. (b) The relationship between the tolerance factor and polarization in double perovskites (black open circles) and DAPs (filled markers). In (b), the polarization of perovskites is plotted in terms of $\mathcal{F}(1-\overline t) \Delta t$, where $\mathcal{F}(x)= c_1x^2+c_2x$. Note that while a single line is shown, APs and perovskites in principle obeydifferent functions with different coefficients. For antiperovskites, $c_1=2.1\times10^4$~$\mu$C/cm$^2$ and $c_2=6.9\times 10^2$~$\mu$C/cm$^2$. Data for perovskites is taken from \cite{mulder2013turning}.}
\end{figure*}

In \cite{mulder2013turning}, Mulder et al. used group theoretical analysis and first-principles calculations to determine the relationship between tolerance factor and polarization in AA'B$_2$O$_6$ double perovskites. They suggested that polarization is a simple function of tolerance factors $t_{ABO_3}$ and $t_{A'BO_3}$:
\begin{equation}
p = \Delta t(c_1(1-\overline t)^2+c_2(1-\overline t)),     \label{eq:Pvst_p}
\end{equation}
where $\Delta t$ and $\overline t$ are the difference and average of the tolerance factors, respectively: 
\begin{equation}
    \Delta t=t_{ABO_3}-t_{A'BO_3}
\end{equation} 
\begin{equation}
\overline{t} = \frac{t_{ABO_3}+t_{A'BO_3}}{2}    
\end{equation}
If $\overline t$ is close to one, the polarization can be approximated as:
\begin{equation}
p \propto \Delta t(1-\overline t).
\end{equation}
This formula is based on simple observations - mode amplitudes of the octahedral rotations ($Q_{M_2^+}$ and $Q_{R_5^-}$) and the A-site displacement ($Q_{X_5^-}$) are proportional to $(1-t_p)$; the cation ordering ($Q_{X_3^-}$) effect on creating a polarization is proportional to the difference in amplitudes between two $X_5^-$ modes, i.e., $\Delta t_p$. Eq.~(\ref{eq:Pvst_p}) offers a straightforward rule for designing double perovskite ferroelectrics with high net polarization. It is natural to ask whether the same formula applies to DAPs. We first follow the procedure developed in \cite{mulder2013turning} to determine the relation between mode amplitudes and tolerance factor $t_a$. As shown in Fig. \ref{fig:Pol_Mode_vs_tol} (a), the equation
\begin{equation}
\sqrt{\overline Q_{M_2^+} \overline Q_{R_5^-}} \propto (1 - \overline{t}_p),     \label{eq::Avrg_Mode_Amp}
\end{equation}
where $\overline Q_{\{M,R\}}$ is defined as
\begin{equation*}
\overline Q_{\{M_2^+,R_5^-\}} = \frac{Q_{\{M_2^+,R_5^-\}} + Q'_{\{M_2^+,R_5^-\}}}{2},     
\end{equation*}
is a good approximation for tetrel DAPs. $Q_{\{M_2^+,R_5^-\}}$ and $Q'_{\{M_2^+,R_5^-\}}$ correspond to mode amplitudes in relaxed APs, calculated standard group theory tools as  implemented in \texttt{ISODISTORT} \cite{stokesisodistort, stokes2006isodisplace}. \texttt{ISODISTORT} is a web-based program that projects the distortion onto irreps of the space group by taking the difference of a parent and a low-symmetry (distorted) phase. A strong correlation between mode amplitudes and tolerance factors is observed in DAPs, similar to in double perovskites \cite{mulder2013turning}, as shown in Fig.~\ref{fig:Pol_Mode_vs_tol}(a). Even though the data for double perovskites and DAPs don't fall on the same line, due to the difference in the nature of chemical bonding in both families of compounds, the DAPs too follow a straight line with a negative slope. 
In Fig.~\ref{fig:Pol_Mode_vs_tol}(b), we plot the polarization of DAPs using  $\mathcal{F}(1-\overline t)\Delta t=\Delta t(c_1(1-\overline t)^2+c_2(1-\overline t))$ as the x axis, and superimpose it with the data on double perovskites from Ref.~\cite{mulder2013turning}. We find that Eqn. (\ref{eq:Pvst_p}) is valid in DAPs as much as it is double perovskites. Thus, the design principles of Ref.~\cite{mulder2013turning} apply to DAPs as well: a DAP with a small average tolerance factor and large tolerance factor difference yields a large polarization.

\subsubsection{Biaxial Strain to Stabilize Ferroelectricity}\label{Sec:Strain}
\begin{figure*}[bht]
\includegraphics[height = 6.5cm]{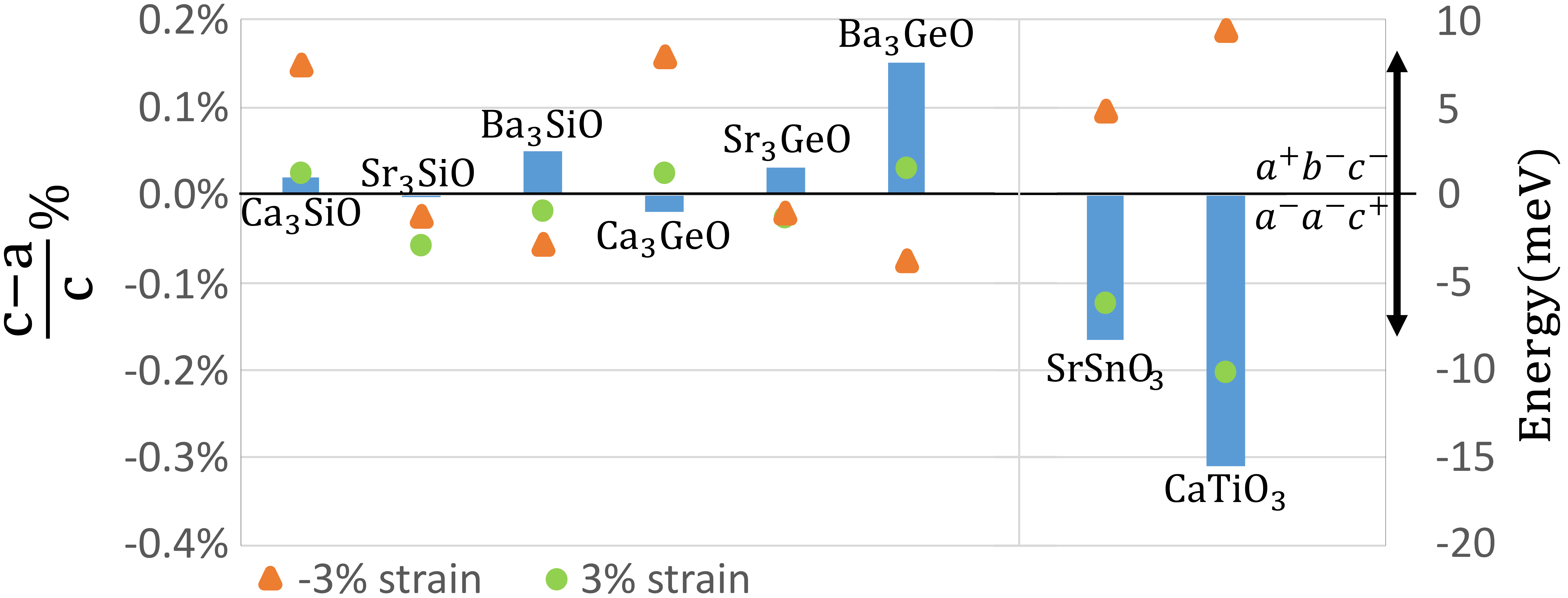}
\caption{\label{fig::strain}Strain effect on the octahedral rotations in APs. The blue bars show the difference between the in-plane and out-of-plane lattice constants ($(c-a)/c$) in the Pbnm phase without strain. The energies of $a^-a^-c^+$ pattern relative to $a^+b^-c^-$ pattern is represented by orange triangles for $-3\%$ strain, and by green circles for $3\%$ strain.}
\end{figure*}

A common approach to engineer the properties of perovskites, including ferroelectricity, is biaxial strain applied through a substrate\cite{schlom2007strain,pertsev1998effect}. In this subsection, we study how biaxial strain affects the octahedral rotations in APs and DAPs, and show that it may provide a promising avenue to stabilize ferroelectricity. Previous studies indicate that in bulk perovskites, the sign of biaxial strain (tensile vs. compressive) can determine whether the $Pbnm$ ($a^-a^-c^+$) or the $P2_1/m$ ($a^+b^-c^-$) structure is more stable \cite{eklund2009strain, wang2018engineering, lu2018tunable, sclauzero2015structural, Saha2024}. This is an important distinction because only the former structure allows hybrid-improper ferroelectricity. 
In order to see if such an effect exists in APs as well, we perform DFT calculations with fixed biaxial strain boundary conditions, where the in-plane lattice parameters are fixed, but the out-of-plane lattice parameter, along with the internal coordinates of atoms, are allowed to relax to minimize energy. Our results, which we present in Fig.~\ref{fig::strain} and discuss below, indicate that there is a weaker coupling between strain and the axes of the octahedral rotations in APs compared to perovskites, which can nevertheless be used to drive these compounds to regimes that favor ferroelectricity.

For simplicity, we only report the results for $-3\%$ and $+3\%$ strain, which are realistic but large values that are sufficient to illustrate the overall trend. Ba$_3$SiO, Sr$_3$SiO and Sr$_3$GeO prefer $Pbnm$ structure, while Ca$_3$SiO and Ca$_3$GeO prefer $P2_1/m$ for both $-3\%$ and $+3\%$ strain. This relatively weak strain effect is a result of  c/a ratio in their orthorhombic phases, which is close to one in APs. We show this quantity as the blue bars in Fig.~\ref{fig::strain}, and for comparison, also show the results for perovskites SrSnO$_3$ and CaTiO$_3$, which have both smaller c/a ratios, and higher sensitivity to strain. 
Ba$_3$GeO is the only AP we studied whose ground state structure is switched by strain. In this compound, compressive (tensile) strain stabilizes the $Pbnm$ ($P2_1/m$) structure. 

\begin{figure}[bht]
\includegraphics[height = 5.2cm]{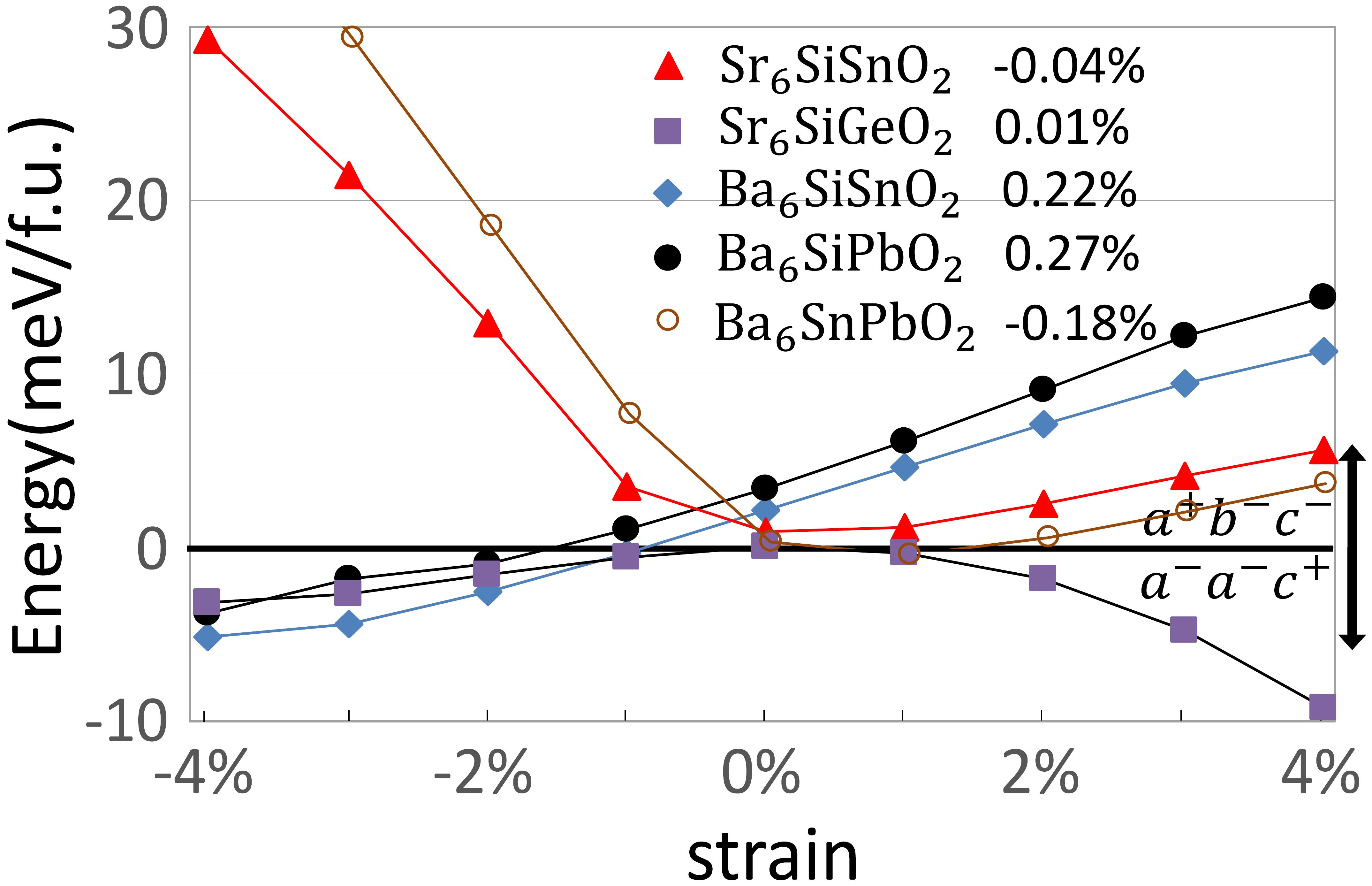}
\caption{\label{fig::strain_2}Energy of $a^-a^-c^+$ tilt systems with respect to $a^+b^-c^-$ systems in DAPs as a function of biaxial strain. The  $(c-a)/c$ values in percent in the $a^-a^-c^+$ phase is shown in the legend.}
\end{figure}

The difference between the two strain orientations becomes more important in DAPs, as $a^+b^-c^-$ tilting yields a non-polar phase. Interestingly, the DAPs show very different trends. We show our results in DAPs in Fig.~\ref{fig::strain_2}. Barium and strontium DAPs display different strain dependences. For barium DAPs, compressive strain stabilizes the $a^-a^-c^+$ system, while tensile strain stabilizes the $a^+b^-c^-$ system. For strontium DAPs, however, $a^+b^-c^-$ is stabilized by both compressive and tensile strain. Compared to tensile strain, the compressive strain raises the energy of $a^-a^-c^+$ even more dramatically. This, however, is not a simple result of the cation only, because   Ba$_6$SnPbO$_2$ displays a trend similar to that in the strontium compounds. To explain the two distinct trends, we again relate the strain effect to c/a ratio. In the legend of Fig. \ref{fig::strain_2}, we report the $c/a$ ratios of DAPs, which explains this trend. If $c/a$ is larger than one, the compressive strain  prefers $a^-a^-c^+$ rotations. On the other hand, if $c/a$ is smaller than one, the compressive strain prefers $a^+b^-c^-$ rotations. Sr$_6$SiGeO$_2$ is the only exception to this empirical observation, and it favors $a^-a^-c^+$ under any strain.
While there seems to be no obvious reason for this exception based on the structural parameters we discussed so far, it is likely that an interplay of biquadratic (4th order) terms with higher order terms is responsible for it. 


\subsection{Electronic Structure and Covalent Bonding in APs}
So far, we explored the structural properties of APs and DAPs. APs have also been studied for their interesting electronic properties, from exotic superconductivity \cite{oudah2016superconductivity,oudah2019evolution} to topological phases \cite{hsieh2014topological,bilal2021dft}. Interplay of ferroelectricity with the electronic structure provides interesting opportunities for controlling these phases \cite{liu2016strain,yoshimi2025emergence,novak2013unusual}. While an extensive study of these phenomena in APs and DAPs is beyond the scope of this study, in this subsection we illustrate some key features of their electronic structures with a focus on covalent bonding. 

\subsubsection{Review of Band Structure of APs}
There are a number of studies that focus on the electronic structure of tetrel oxide APs \cite{kariyado2011three,oudah2016superconductivity,ochi2019comparative,hassan2018structural}. Many of these compounds exhibit narrow gap semiconducting or semimetallic behavior. Due to their small band gaps and thus easy band inversion, studies of topology in AP oxides yield fruitful results in this area: For example, Sr$_3$PbO is a potential crystalline topological insulator, protected by mirror reflection about (100) plane \cite{hsieh2014topological}. 
Furthermore, due to the mixing of Sr's 4d and Sn's 5p orbitals, the topological superconducting phases are predicted in Sr$_{3-x}$SnO with appropriate Sr deficiency \cite{oudah2016superconductivity, oudah2019evolution}. Weak antilocalization is found when the Fermi level is around the Dirac nodes due to spin-momentum entanglement in this compound as well \cite{nakamura2020robust}. 

\begin{figure}[b]
\centering
\includegraphics[width=1.05\linewidth]{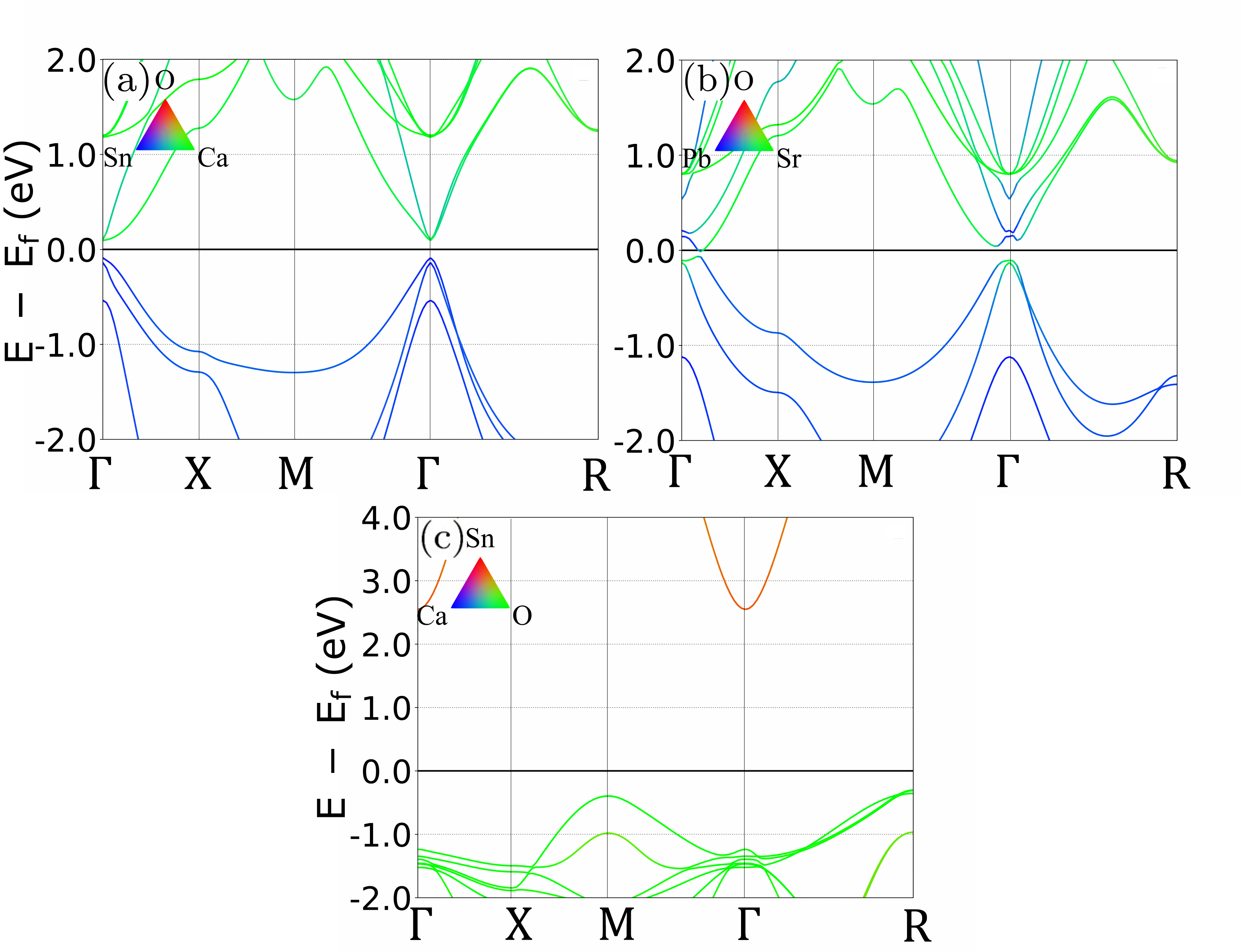}
\caption{The atom-projected band structures for (a) Ca$_3$SnO, (b) Sr$_3$PbO, (c) CaSnO$_3$. The calculations are performed using the Heyd, Scuseria, and Ernzerhof (HSE) hybrid functional which is necessary for capturing the details of the bandstructure \cite{heyd2003hybrid}. Spin-orbit coupling is also taken into account. } \label{fig:Band}
\end{figure}

Our calculations using HSE hybrid functional \cite{heyd2003hybrid} and spin orbit coupling confirm this result: The nontrivial topological phase is realized in Sr$_3$PbO by the band inversion between Sr-d orbitals and Pb-p orbitals at $\Gamma$ point, and a small gap opening along $\Gamma-X$ \cite{hsieh2014topological}, as shown in Fig. \ref{fig:Band}(b). Another tetrel oxide AP, Ca$_3$SnO, is topologically trivial, with Ca-d orbitals lying above Sn-p orbitals, creating a wider band gap (Fig. \ref{fig:Band}(a)). Even in this case, however, the gap is well below that of, for example CaSnO$_3$, shown in Fig.~\ref{fig:Band}(c). 

The prediction of a topological crystalline insulating phase  makes studying ferroelectricity in this family of compounds particularly interesting, since the surface states in this type of topological phases can in principle be controlled by the ferroelectric order, as discussed for CsPbI$_3$ \cite{liu2016strain}.

\begin{figure*}[bht]
\includegraphics[width=0.94\linewidth]{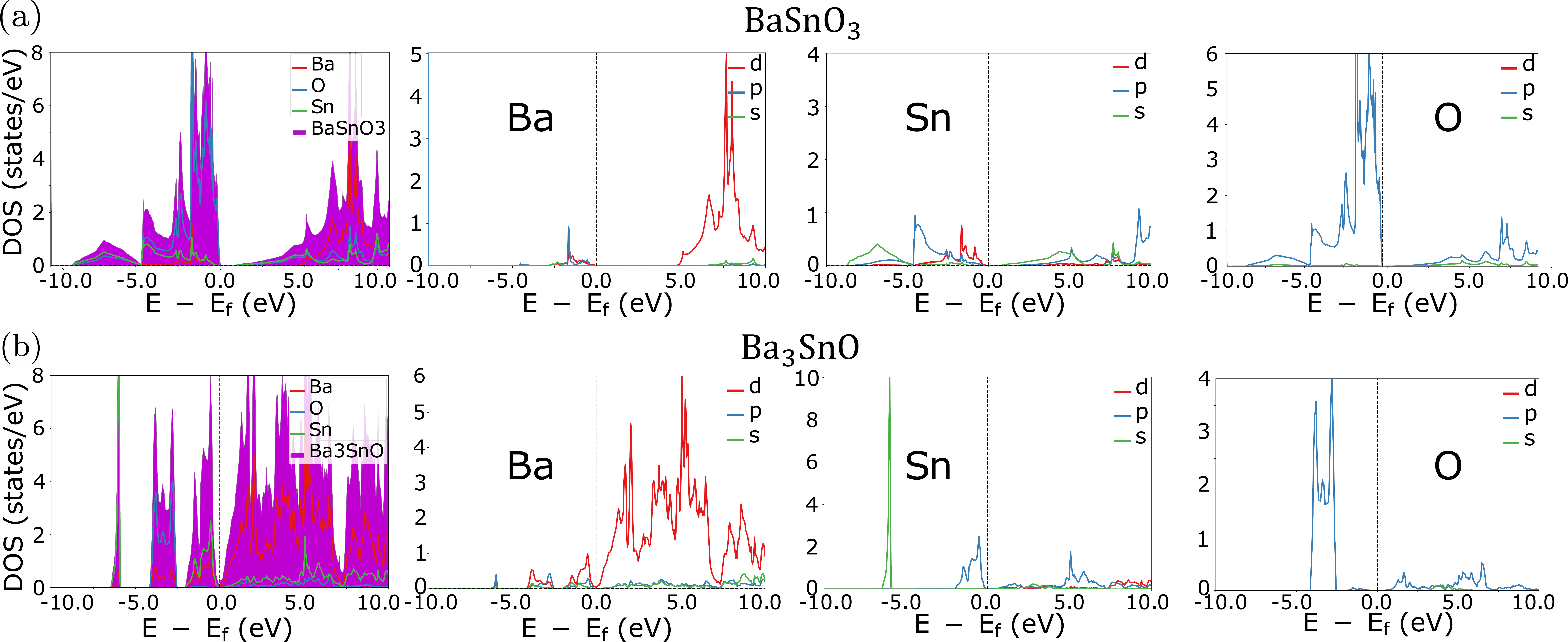}
\caption{\label{fig:DOS_all}Atomic and orbital resolved DOS plots for (a) BaSnO$_3$, and (b) Ba$_3$SnO.}
\end{figure*}

Apart from the band gap and the character of the bands near the Fermi level, the energy resolved DOS at a larger scale of APs is completely different from that of perovskites as well. In Fig.~\ref{fig:DOS_all}, we compare the densities of states of BaSnO$_3$ and Ba$_3$SnO. While the alkaline earth DOS is often irrelevant to properties near the Fermi level in perovskites, the opposite is true in antiperovskites and Ba$_3$SnO conduction band is dominated by Ba character. On the other hand, there are no oxygen states near the Fermi level. These differences are reflected in the covalent bonding in APs as well, as discussed below in this section. 

\subsection{Bandgap of DAPs}

In order to evaluate the potential of electric field control of polarization in DAPs, we performed bandstructure calculations for all DAPs we considered as well. The bandstructures are shown in Supplementary Fig. S3 \cite{supplement}. These calculations are performed using the PBEsol exchange-correlation functional, and hence provide a lower estimate for the actual bandgap. In the paraelectric phase without octahedral rotations, all DAPs are (semi-)metallic, with overlapping conduction and valence bands near the $\Gamma$ point. The octahedral rotations tend to narrow the bands, and hence open a band gap in some DAPs, as shown in Table~\ref{tab:dapsgap}. For most ferroelectric DAPs (i.e. DAPs with a polar crystal structure driven by the hybrid-improper mechanism), the band gap remains narrow enough that the potential of these materials serving as good ferroelectrics is hampered. Nevertheless, the `ferroelectric-like' structures we discussed are relevant, because there is an increasing number of studies that theoretically predict or experimentally observe, for example, the coexistence of free carriers and  ferroelectricity \cite{shi2013ferroelectric, benedek2016ferroelectric}, switching of polar order in a metal \cite{fei_ferroelectric_2018}, and even the coexistence of superconductivity with ferroelectricity \cite{jindal_coupled_2023}. 

\begin{table}
\begin{tabular}{c c}
\hline
Compound & Band gap (eV) \\ \hline
  Ca$_6$SiGeO$_2$ & 0.14 \\
  Sr$_6$SiGeO$_2$ & 0.35 \\
  Sr$_6$SiSnO$_2$ & 0.04 \\
  Sr$_6$GeSnO$_2$ & 0.11 \\
  Ba$_6$SiGeO$_2$ & 0.34\\
\end{tabular}
\caption{Band gaps of insulating ferroelectric DAPs obtained using PBEsol exchange-correlation functional. Due to the underestimation of bandgaps by generalized gradient approximations such as PBEsol, these numbers serve only as a lower limit to the actual band gap in these materials. The k-dependent bandstructures of these compounds are reported in Supplementary Figure S4 \cite{supplement}.}
\label{tab:dapsgap}
\end{table}

\subsubsection{Covalent Bonding in APs}
The discussion in the previous sections about octahedral rotations, tolerance factor, and cation ordering is largely based on the ionic character of APs. However, even in highly ionic perovskites, the covalency and hybridization give rise to many interesting phenomena, such as proper ferroelectricity (e.g. in BaTiO$_3$ \cite{cohen1992origin}) or the magnetoelectric effect (e.g. in EuTiO$_3$ \cite{birol_origin_2013}). Covalent interactions in primarily ionic perovskites also introduce corrections to energetics of oxygen octahedral rotations \cite{woodward1997octahedralII, cammarata2014covalent}. For example, in YAlO$_3$, the preference for $a^-a^-c^+$ disappears as the Y-O covalent interaction is turned off \cite{woodward1997octahedralII}. In this section, we focus on the bonding in APs to develop a better understanding of the interplay of electronic and crystal structure in these compounds. 

The DOS shown in Fig.~\ref{fig:DOS_all}(b) shows signature of strong inter-ionic hybridization in Ba$_3$SnO AP. The nominally empty d-orbitals of Ba form bands that also display O character above the Fermi level, indicating some covalency between these ions. Similarly, the occupied Sn-s bands below the Fermi level mix strongly with the Ba-d orbitals. This implies that both A-X and B-X covalent interactions could be important in APs, unlike oxide perovskites, where the A-site ions' hybridization with the X (oxygen) ions is often negligible. 

We now use the crystal orbital Hamilton population (COHP) approach to elucidate the covalent bonding trends in APs.  COHP, a common first principles chemistry tool, allows distinguishing bonding and antibonding contributions as a function of energy \cite{dronskowski1993crystal}. 
%
%
A negative COHP between two orbitals corresponds to bonding interactions, whereas positive COHP corresponds to antibonding interactions. The integral of energy resolved COHP up to the Fermi level (integrated COHP, or ICOHP) yields a number with units of energy that can be used to evaluate the strength of covalent bonding between a pair of ions or orbitals. While  comparison of ICOHP between different systems is not perfectly quantitative because of reasons including the incompleteness of basis sets, the relative percent contribution of ICOHP of different bonds provides an estimate of relative bond strengths \cite{dronskowski1993crystal, steinberg2012identifying,steinberg2015electron}.

\begin{figure}[bht]
\includegraphics[width=0.9\linewidth]{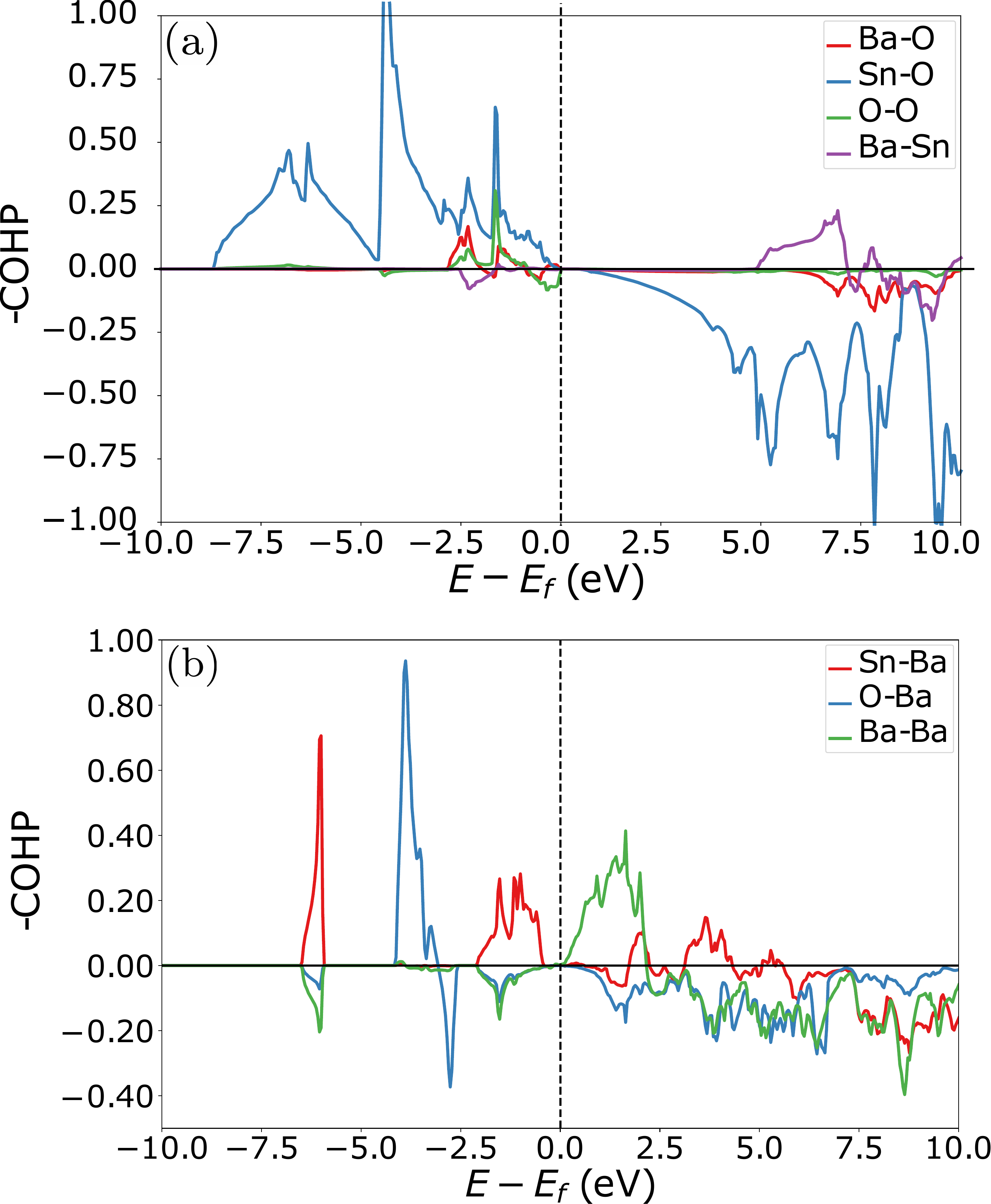}
\caption{\label{fig::COHP}-COHP between nearest neighbors as a function of energy for (a) BaSnO$_3$ and (b) Ba$_3$SnO.}
\end{figure}

In Fig.~\ref{fig::COHP}(a) and (b), we show the  -COHP (instead of COHP, by convention) plots of atomic pairs in BaSnO$_3$ and Ba$_3$SnO, respectively. In these plots, positive values below the Fermi level shows covalent bonding. In the perovskite BaSnO$_3$ , Sn-O bonds contribute the most to the total bonding interactions, while only a small portion originates from Ba-O and O-O bonds near the Fermi level. This is also evidenced by the -ICOHP values in Table S1: -ICOHP sums to 13.76 eV/cell for Sn-O bonds, whereas to 0.86 eV/cell for Ba-O bonds. States above the Fermi level between each pair are highly anti-bonding. While there is some Ba-O bonding and Ba-Sn antibonding character below the Fermi level, these are negligible. In the AP, however, the -COHP curves are very different and more complex:  Unlike the dominance of a single type of (B-O) bonds in perovskites, multiple atomic pairs display significant -COHP below the Fermi level. There is significant bonding between Sn-s and p orbitals  with Ba-d (hybridized with s and p), ranging from -6.5 to -5.9 eV and -2.1 to -0.4 eV, respectively (Table S2). These bonds add up to an ICOHP of 4.43 eV/cell. Ba-O bonding states range from -4.2 to -3.1 eV, with some anti-bonding interactions between -3.1 to -2.6 eV, and a ccumulative -ICOHP value of 1.23 eV/cell. 
Interestingly,  Fig. \ref{fig::COHP}(b) shows that contribution from Ba-Ba pairs to -COHP is negative, indicating anti-bonding interactions, which is the opposite of what happens in perovskites, where there is often some  bonding between ions occupying the face-center sites, as is the case with oxygens in SrSnO$_3$ (Fig.~\ref{fig::COHP}(a)). 
The sum of Ba-Ba -ICOHP in Ba$_3$SnO is -1.66 eV/cell. 

The percentage contribution of different bonds in perovskites and antiperovskites is presented in Fig.~\ref{fig:histogram}. (The data for this figure is shown in supplementary Tables S1 and S2 \cite{supplement}.) There are three main differences between APs and perovskites. First, the largest bonding interaction in APs comes from A-X pairs in APs, instead of B-X as in perovskites. This is partly due to the charge state of the tetrel elements in APs, as a result of which their highest occupied band close to the Fermi level is high enough in energy that it can mix with the unoccupied alkaline-earth d bands. However, even though the A-X bonds are the strongest in terms of -ICOHP, other bonds in APs contribute about 45\% on average to the total -ICOHP. On the other hand, the B-X bonds contribute more than 80\% of the total ICOHP in perovskites by themselves. This means that the APs are electronically much more mixed than perovskites, with large covalent mixing between all different ions. Second, unlike perovskites, where few anti-bonding states are occupied, all APs we considered exhibit by sizable anti-bonding ICOHP below the Fermi level. Third, in APs, the ICOHP (bonding or antibonding) between electronic states of different X atoms is sizable, especially in barium and calcium APs. In fact, the X-X interactions can be even stronger than B-X interactions. The average contribution is $15.7\%$ for APs, compared with $4.4\%$ for perovskites. As a result, while the O-O bonds are often ignored in perovskites, the X-X bonds need to be considered in antiperovskites, and simplified models that aim to reproduce their bandstructures.

\begin{figure*}[bht]
\includegraphics[height = 11cm]{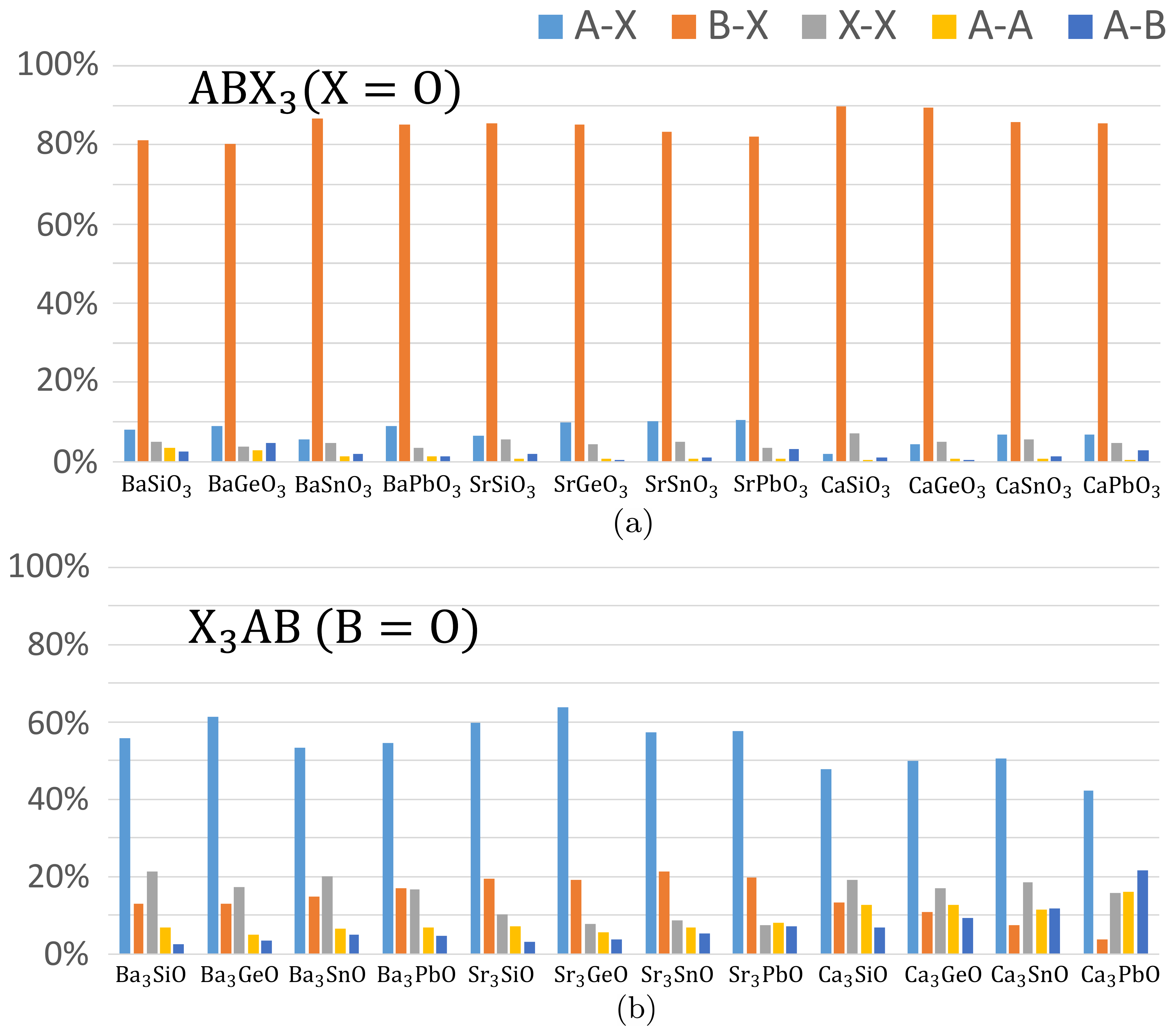}
\caption{\label{fig:histogram}Contribution of different bonds to total ICOHP in (a) perovskites, and (b) APs. The data in these figures are also presented in Supplementary Tables SII and SIII. }
\end{figure*}

\section{Conclusions}
In this study, we performed a comprehensive theoretical analysis of the crystal structure of X$_3$AB antiperovskites, where X is an alkaline earth metal (Ca, Sr, Ba), A is a tetrel element (Si, Ge, Sn, Pb), and B is oxygen. Our first principles DFT results confirm the earlier experimental observations of oxygen octahedral rotations in these systems, and show that all the experimentally observed phases can be stabilized in DFT calculations. The energetics predict that lower temperature phases orthorhombic phases should be present in some of the AP systems, which can be observed by experiments performed at lower temperatures than those used so far. 

Just as the octahedral rotation phases in APs follow tolerance factor arguments originally developed for perovskites, heterostructures of APs (the so-called double antiperovskites) can be studied using the theoretical machinery developed to predict emergence of hybrid-improper ferroelectricity in double perovskites. By performing such an analysis, we showed that it is indeed possible to design DAP structures, which naturally form the necessary cation order to induce hybrid improper ferroelectricity. While these systems have either semi-metallic behavior or too small gaps to be useful in applications that require strong ferroelectrics with reliable switching and low loss, these systems with their polar crystal structures and strong mixing between different bands near the Fermi level are useful candidates to study the coupling between crystal structure and topological or superconducting states predicted in AP systems. 

Finally, quantitative analysis of covalent bonding in APs, as studied by calculating both the energy resolved COHP and the ICOHP from first principles, show that the inter-ionic mixing is much more complicated in APs, compared to their perovskite counterparts with the same elements. While these systems are still highly ionic, the richness of features in the COHP plots explains the variety of electronic phenomena predicted or observed in these systems.


\acknowledgements
This work is supported by the Office of Naval Research Grants No. N00014-20-1-2361 and N00014-24-1-2082.

\bibliography{reference}

\end{document}